\documentclass[12pt]{article}

\ifx\pdfoutput\undefined
\usepackage[dvips,bookmarks]{hyperref}
\else
\usepackage{hyperref}
\fi
\hypersetup{colorlinks=false,bookmarksopen,bookmarksnumbered,citecolor=blue,
   pdfstartview=FitH}

\usepackage{latexsym}

\usepackage{amssymb,amsfonts,amsmath}
\usepackage{graphicx} 
\usepackage{indentfirst}

\usepackage{tensor}

 \usepackage{bbm}

\parskip=\medskipamount

\usepackage[mathscr]{euscript} 

\newcommand {\cD}{{\cal D}}

\newcommand {\cF}{{\cal F}}

\newcommand {\cJ}{{\cal J}}

\newcommand {\cL}{{\cal L}}
\newcommand {\cM}{{\cal M}}
\newcommand {\cN}{{\cal N}}

\newcommand {\cV}{{\cal V}}
\newcommand {\cW}{{\cal W}}

\newcommand {\cY}{{\cal Y}}
\newcommand {\cZ}{{\cal Z}}
\newcommand{\dsC}{{\mathbb C}}

\newcommand{\dsJ}{{\mathbb J}}
\newcommand{\dsL}{{\mathbb L}}

\newcommand{\dsR}{{\mathbb R}}
\newcommand{\dsW}{{\mathbb W}}

\def\ds1{\ensuremath{\mathbbm{1}}}
\def\a{\alpha}
\def\b{\beta}

\def\d{\delta}

\def\g{\gamma}

\def\l{\lambda}
\def\m{\mu}

\def\q{\theta}

\def\s{\sigma}

\def\x{\xi}
\def\z{\zeta}
\def\D{\Delta}
\def\F{\Phi}

\def\O{\Omega}

\def\S{\Sigma}

\newcommand{\eps}{\varepsilon}
\newcommand{\vf}{\varphi}
\newcommand{\da}{{\dot{\alpha}}}
\newcommand{\db}{{\dot{\beta}}}
\newcommand{\dg}{{\dot{\gamma}}}
\newcommand{\dd}{{\dot{\delta}}}

\newcommand{\Db}{{\bar D}}

\newcommand{\Yb}{{\bar Y}}
\newcommand{\Wb}{{\bar W}}
\newcommand{\cDb}{{\bar \cD}}

\newcommand{\cWb}{{\bar \cW}}
\newcommand{\cYb}{{\bar \cY}}
\newcommand{\cZb}{{\bar \cZ}}
\newcommand{\Sb}{{\bar S}}
\newcommand{\cFb}{{\bar \cF}}

\newcommand{\Mb}{\bar{M}}

\newcommand{\1}{{\underline{1}}}
\newcommand{\2}{{\underline{2}}}

\newcommand{\bm}[1]{\boldsymbol{#1}}
\newcommand{\be}{\begin{equation}}
\newcommand{\ee}{\end{equation}}
\newcommand{\bea}{\begin{eqnarray}}
\newcommand{\eea}{\end{eqnarray}}
\newcommand{\non}{\nonumber}
\newcommand{\bem}{\begin{pmatrix}}
\newcommand{\eem}{\end{pmatrix}}
\def\tr{{\rm tr}}

\def\ad{{\rm Ad}} 

\def\half{\ensuremath{\frac{1}{2}}}

\def\const {{\rm const}}

\def\cc {{\rm c.c.}}

\def \const {{\rm const}}

\def\a{\alpha}

\def\b{\beta}

\def\d{\delta}

\def\g{\gamma}

\def\l{\lambda}
\def\m{\mu}

\def\o{\omega}

\def\q{\theta}

\def\s{\sigma}

\def\x{\xi}
\def\z{\zeta}
\def\D{\Delta}
\def\F{\Phi}

\def\O{\Omega}

\def\S{\Sigma}

\def\tr{{\rm tr}}
\def\rd{{\rm d}}
\def\ri{{\rm i}}
\def\re{{\rm e}}

\newcommand{\ve}{\varepsilon}                            
\newcommand{\cDB}{{\bar\cD}}                            

\newcommand{\pa}{\partial}                           
\newcommand{\hf}{\frac12}
\def\double #1{#1{\hbox{\kern-2pt $#1$}}}

\newcommand{\bsubeq}{\begin{subequations}}
\newcommand{\esubeq}{\end{subequations}}

\newcommand{\Gb}{\bar{G}}
\newcommand{\Fb}{\bar{F}}

\newcommand{\cMb}{\bar{\cM}}
\newcommand{\Phib}{\bar{\Phi}}

\begin{document}
\begin{titlepage}
\begin{flushright}
October, 2011\\
\end{flushright}

\begin{center}
{\Large \bf 
Vector-tensor supermultiplets in AdS \\and supergravity}
\end{center}

\begin{center}

{\bf
Sergei M. Kuzenko\footnote{sergei.kuzenko@uwa.edu.au}
and
Joseph Novak\footnote{joseph.novak@uwa.edu.au}
} \\
\vspace{5mm}

\footnotesize{
{\it School of Physics M013, The University of Western Australia\\
35 Stirling Highway, Crawley W.A. 6009, Australia}}  
~\\
\vspace{2mm}

\end{center}

\begin{abstract}
\baselineskip=14pt
In 
$\cN=2$ Poincar\'e supersymmetry in four space-time dimensions, 
there exist off-shell supermultiplets with intrinsic central charge, including 
the important examples of the Fayet-Sohnius hypermultiplet, 
the linear and the nonlinear vector-tensor (VT) multiplets. 
One can also define similar supermultiplets in the context of $\cN=2$ anti-de Sitter (AdS) supersymmetry, 
although the origin of the central charge becomes somewhat obscure. In this paper we develop a general setting 
for $\cN=2$ AdS supersymmetric theories with central charge.  We formulate a 
supersymmetric 
action principle in $\cN=2$ AdS superspace and then reformulate it in terms of $\cN=1$ superfields.
We prove that $\cN=2$ AdS supersymmetry does not allow existence of a linear VT multiplet. For the nonlinear 
VT multiplet, we derive consistent superfield constraints in the presence of  any number of 
$\cN=2$ Yang-Mills vector multiplets,
give the supersymmetric action and elaborate on the $\cN=1$ superfield and component descriptions 
of the  theory. Our description of the nonlinear VT multiplet in AdS is then lifted to $\cN=2$ supergravity. 
Moreover, we derive consistent superfield constraints and Lagrangian that describe the linear
VT multiplet in $\cN=2$ supergravity in the presence of  two vector multiplets, 
one of which gauges the central charge.
These supergravity constructions
thus provide the first superspace formulation for the component results  derived in arXiv:hep-th/9710212.
We also construct higher-derivative couplings 
of the VT multiplet to any number of $\cN=2$ tensor multiplets.
\end{abstract}

\vfill
\end{titlepage}

\newpage
\renewcommand{\thefootnote}{\arabic{footnote}}
\setcounter{footnote}{0}

\tableofcontents{}
\vspace{1cm}
\bigskip\hrule

\section{Introduction}
\setcounter{equation}{0}

The vector-tensor (VT) supermultiplet is a dual version of the  Abelian $\cN=2$ vector multiplet
in four dimensions,   obtained by dualizing 
one of the two physical scalars (belonging to the vector multiplet) into a gauge two-form.
The auxiliary fields of these multiplets also differ, namely: a real isotriplet in the vector multiplet case, 
and a real scalar in the dual version.
In contrast to the vector multiplet, the VT multiplet has an off-shell central charge, 
similar to the Fayet-Sohnius hypermultiplet \cite{Fayet,Sohnius}. 

The history of the VT multiplet is quite interesting.
It was discovered by Sohnius, Stelle and West \cite{SSW}  in 1980 (see \cite{SSW2} for a review) as 
a spin-off of their attempts to construct an off-shell formulation for extended supersymmetric gauge theories.
Soon after,  it was shown by Milewski \cite{Milewski}  (see also \cite{Milewski2} for a review)
that this multiplet has a simple structure from the point of view 
of $\cN=1$ supersymmetry. Specifically, its action functional in $\cN=1$ superspace is the  sum of those
describing $\cN=1$ vector and tensor multiplets
\bea
S_{\text{VT}} =  \hf  \int {\rm d}^4  x \rd^2 \q  \,
W^\a W_\a 
- \hf \int {\rm d}^4 x \rd^4 \q \, G^2~.
\label{1.1}
\eea
Here $W_\a$ is the chiral field strength of the $\cN=1$ vector multiplet, 
\bea
\bar D_\da W_\a =0~, \qquad D^\a W_\a = \bar D_\da \bar W^\da~, 
\eea 
while $G$ is the real linear field strength of the $\cN=1$ tensor multiplet \cite{Siegel}, 
\bea
\bar G = G~, \qquad \bar D^2 G=0~.
\eea
Then, the VT multiplet was completely forgotten for over a decade.

Research on the VT multiplet experienced a  renaissance  in  the year 1995 when 
de Wit, Kaplunovsky, Louis and L\"ust  \cite{deWKLL} realized that this multiplet describes 
the dilaton-axion complex in heterotic $\cN=2$ four-dimensional supersymmetric string 
vacua.\footnote{From a historical point of view, it is of interest to mention
that Ref \cite{deWKLL} in fact announced the discovery of the  VT multiplet, in spite of the
existence of the original \cite{SSW,Milewski} and review \cite{SSW2,Milewski2} papers on the VT multiplet
published in the early 1980s.
It thus appears that this multiplet had been completely forgotten by mid-1990s.} 
This work triggered numerous studies of the VT multiplet and its Chern-Simons couplings in the component
field approach (both in rigid supersymmetry and supergravity using the 
superconformal tensor calculus)  \cite{Claus1,Claus2,Claus3}, as well as 
in the framework of conventional $\cN=2$ superspace \cite{HOW,GHH,BHO}
and $\cN=2$ harmonic superspace \cite{DKT,DK,IS,DIKST}
(see also \cite{DT}).
In particular, it was found that besides the original `linear' VT multiplet \cite{SSW}, there exists 
a `nonlinear' VT multiplet \cite{Claus1,Claus2}. The difference between these inequivalent 
realizations is quite transparent in $\cN=2$ superspace, and here we would like to discuss this issue in some detail. 

Let us consider $\cN=2$ central charge superspace \cite{Sohnius}.  
The spinor derivatives with central charge, $\D$, form the algebra
\begin{subequations}\label{1.4}
\begin{align}
\{ D_\a^i, D_\b^j \} &= \phantom{-}2 \eps_{\a\b} \eps^{ij} \D  ~,\\
\{ \Db^\da_i, \Db^\db_j \} &= - 2 \eps^{\da\db} \eps_{ij} \D ~, \\
\{ D_\a^i , \Db^\db_j \} &= -2 \ri \d^i_j \partial_\a{}^\db ~.
\end{align}
\end{subequations}
Following \cite{GHH,DKT,BHO}, 
the  {\it linear VT multiplet} can be described by a real superfield, $L$,  constrained by 
\begin{subequations}\label{1.5}
\bea D^{ij} L &=&    0~, \\
D_\a^{(i} \Db_\da^{j)} L &=& 0 ~,
\eea
\end{subequations}
where we have denoted $D^{ij}  :=  D^{\a (i} D_{\a}^{j)}$ and 
$\bar D_{ij} := \bar D_{\da (i} \bar D^\da_{j)}$. 
The multiplet is on-shell,  $\Box L =0$, if $\D L =0$.
Following \cite{DK}, the {\it nonlinear VT multiplet} is described by a real superfield $\mathbb L$ 
subject  to the constraints
\begin{subequations}\label{1.6}
\begin{align} D^{ij} \dsL &= 2 \kappa D^i \dsL D^j \dsL - \kappa \Db^i \dsL \Db^j \dsL ~, \\
D_\a^{(i} \Db_\da^{j)} \dsL &= \kappa D_\a^{(i} \dsL \Db^{j)}_\da \dsL \ ,
\end{align}
\end{subequations}
where $\kappa$ is a real coupling constant of inverse mass dimension. 
These constraints can be written in an alternative form \cite{IS}
using a new superfield $L = \exp{(- \kappa \dsL)}$. One finds 
\begin{subequations}\label{1.6-mod}
\begin{align} 
D^{ij} L &=- \frac{1}{L} D^i L D^j L + \frac{1}{L} \bar D^i L \bar D^j L~, \\
D_\a^{(i} \bar D_\da^{j)} L &= 0 \ . 
\end{align}
\end{subequations}
We can think of the constraints (\ref{1.6}) as the unique consistent deformation of
(\ref{1.5}), see  \cite{DK} for more details. The two VT multiplets have different Chern-Simons 
couplings to vector multiplets \cite{Claus1,Claus2,DIKST}, 
including the one that gauges the central  charge, and to $\cN=2$ supergravity \cite{Claus3}.

It turns out the constraints for the VT multiplets have an  interesting higher-dimensional origin.
The linear VT multiplet constraints, eq. (\ref{1.5}), can be interpreted as the equations obeyed by 
the gauge-invariant superfield strength of a free on-shell $\cN=1$ vector multiplet in five dimensions \cite{KL}. 
The same constraints also admit a six-dimensional origin \cite{DIKST} in terms of the (1,0) self-dual tensor 
multiplet \cite{BSV,Sokatchev}. The nonlinear VT multiplet constraints, eq. (\ref{1.6-mod}), 
coincide with the equations of motion for the five-dimensional $\cN=1$ supersymmetric U(1) Chern-Simons 
theory \cite{KL}.

One of our goals in this paper is to study VT multiplets and their couplings in 
four-dimensional $\cN=2$ anti-de Sitter (AdS) supersymmetry. 
It is known that $\cN=1, 2$ rigid supersymmetric theories in AdS differ significantly from 
their counterparts defined in Minkowski space 
\cite{BK_AdS_supercurrent,AJKL,FS,BKsigma1,BKsigma2,Jia:2011hw}. 
We therefore may expect nontrivial restrictions to VT multiplet interactions in AdS. 
Indeed, our first observation is that a linear VT multiplet does not exist in AdS.
The simplest way to prove this claim is by using a formulation in $\cN=1$ AdS superspace. 

Let us assume that  there exists an AdS extension of the linear VT multiplet. Then its dynamics can be formulated in 
$\cN=1$ AdS superspace where the linear VT multiplet decomposes into a vector and a tensor multiplet.
The corresponding action should be a minimal AdS extension of (\ref{1.1}), that is 
\bea
S_{\text{LVT}} =  \hf  \int {\rm d}^4 x \rd^4 \q  \, \frac{E}{\m}\,
W^\a W_\a 
- \hf \int {\rm d}^4 x \rd^4 \q \,E \, G^2~,
\label{1.7}
\eea
where $W_\a$ is the covarianltly chiral field strength of the vector multiplet, 
\bea
\bar \nabla_\da W_\a =0~, \qquad \nabla ^\a W_\a = \bar \nabla _\da \bar W^\da~, 
\eea 
and $G$ is the real linear field strength of the tensor multiplet, 
\bea
\bar G = G~, \qquad ( \bar \nabla^2 -4\m) G=0~.
\eea
The tensor multiplet sector of (\ref{1.7}) can be dualized into a covariantly chiral superfield $\F$, $\bar \nabla_\da \F=0$,
 and its conjugate $\bar \F$ \cite{Siegel}. Then the above action turns into 
 \bea
S =  \hf  \int {\rm d}^4  x \rd^4 \q  \, \frac{E}{\m}\,
W^\a W_\a 
+ \hf \int {\rm d}^4 x \rd^4 \q \,E \, (\F +\bar \F)^2~.
\label{1.10}
\eea
On the other hand, the linear VT multiplet should be dual to a free  $\cN=2$ vector multiplet. 
The latter is described in $\cN=1$ AdS superspace by the action (see, e.g., \cite{KT-M-AdS})
\bea
S_{\text{vector}} =  \hf  \int {\rm d}^4  x \rd^4 \q  \, \frac{E}{\m}\,
W^\a W_\a + \int {\rm d}^4 x \rd^4 \q \,E \, \bar \F \F~.
\label{1.11}
\eea
By assumption, the dynamical systems (\ref{1.10}) and (\ref{1.11}) should be equivalent to each other.
However the chiral sectors of (\ref{1.10}) and (\ref{1.11}) are different. 
This means that a free linear VT multiplet does not exist in AdS.\footnote{In supergravity, 
it was shown in \cite{Claus3} that the linear VT multiplet can be consistently defined 
in the presence of an Abelian vector multiplet in addition to the central charge vector multiplet.
In the rigid supersymmetric case, it was demonstrated in \cite{Claus2,DIKST} that 
in the case of the linear VT multiplet with gauged central charge one also needs at least two vector multiples
(one of which is associated with the central charge) 
for ensuring the rigid scale and chiral symmetries of  the action.}

The above example provides enough rationale for studying the VT multiplet and its couplings  in AdS. 
In $\cN=2$ supergravity, on the other hand, 
the VT multiplets and their couplings have been studied only in the component approach \cite{Claus3}. 
Developing a superspace formulation appears to be  highly desirable.

This paper is organized as follows. In section 2  we describe  a general setting 
for $\cN=2$ AdS supersymmetric theories with central charge and formulate 
a supersymmetric action principle in $\cN=2$ AdS superspace. 
In section 3 we derive consistent superfield constraints and a Lagrangian for the VT multiplet 
in $\cN=2$ AdS superspace, 
and then generalize them to include couplings to vector multiplets.  The results of section 3 are 
then reformulated in $\cN=1$ AdS superspace in section 4. Section 5 is devoted to component results. 
Extension of our AdS constructions to supergravity is given in section 6. 
We sketch some interesting generalizations of our results and discuss open problems in section 7.
The main body of the paper is accompanied by four technical appendices. 
Appendix A contains salient facts about $\cN=1$ AdS superspace. Appendix B is devoted to $\cN=2$ 
Killing vector fields. Some aspects of $\cN=2 \to \cN=1$ AdS superspace reduction are discussed in 
Appendix C. Finally Appendix D contains a summary of the  superspace formulation for $\cN=2$ conformal supergravity. Our notation and two-component spinor conventions follow \cite{BK}.

\section{N = 2 AdS supersymmetry and central charge} \label{N2ADS}
\setcounter{equation}{0}

The four-dimensional $\cN=2$ AdS superspace 
$$
{\rm AdS^{4|8} } := \frac{{\rm OSp}(2|4)}{{\rm SO}(3,1) \times {\rm SO} (2)}
$$
can be realized as a maximally symmetric geometry that originates within 
the superspace formulation of $\cN=2$ conformal supergravity developed in  \cite{KLRT-M08}.
Assuming the superspace is parametrized  by local bosonic ($x$) and fermionic ($\q, \bar \q$) 
coordinates  ${\bm z}^{\cM}=(x^{m},\q^{\mu}_{\imath},{\bar \q}_{\dot{\mu}}^{\imath})$
(where $m=0,1,\cdots,3$, $\mu=1,2$, $\dot{\mu}=1,2$ and  $\imath=\1,\2$),
the corresponding covariant derivatives 
\bea
{\cD}_{A} =({ \cD}_{a}, { \cD}_{{\a}}^i, { \cDB}^\da_i)
= {E}_{A}{}^M \pa_M + \hf  {\O}_{A}{}^{ bc} M_{ bc} 
+  \F_{A} {S}^{ij} J_{ij}~, \qquad i,j =\1 , \2
\label{2.1}
\eea
obey the algebra  \cite{KT-M-AdS, KLRT-M08}
\begin{subequations}  \label{AdSD}
\begin{align} \{ \cD_\a^i , \cD_\b^j \} &= 4 S^{ij} M_{\a\b} + 2 \eps_{\a\b} \eps^{ij} S^{kl} J_{kl} \ , 
\qquad \{ \cD_\a^i, \cDb^\db_j\} = - 2 \ri \d^i_j \cD_\a{}^\db ~, \\
[\cD_a , \cD_\b^j] &= \frac{\ri}{2} (\s_a)_{\b\dg} S^{jk} \cDb_k^\dg \ , \qquad  [\cD_a , \cD_b] = - S^2 M_{ab}~,
\end{align}
\end{subequations}
where ${ S}^{ij} $ is a {\it covariantly constant} and {\it constant}
 real isotriplet, ${ S}^{ji} = {S}^{ij}$, 
 $\overline{ { S}^{ij}} = { S}_{ij} =\ve_{ik}\ve_{jl}{ S}^{kl}$, 
and  ${ S}^2 :=  \frac{1}{2} { S}^{ij} { S}_{ij} $. 
The SU(2) generators, $J_{kl}$, act on the spinor covariant derivatives 
by the rule:
\bea
[J_{kl} , { \cD}_\a^i ] = -\hf ( \d^i_k  { \cD}_{\a l} +\d^i_l  { \cD}_{\a k})~.
\eea
This superspace proves to be a conformally flat solution to the equations of motion 
for $\cN=2$ supergravity with a cosmological term \cite{BK_AdS_supercurrent}.

Our goal in this section is to develop a general setting to formulate $\cN=2$ rigid supersymmetric theories 
with an off-shell central charge in the $\cN=2$ AdS superspace introduced. As compared with the 
super-Poincar\'e case, eq. (\ref{1.4}), there appears to be a subtlety: the algebra of  the AdS covariant derivatives
(\ref{AdSD}) cannot be deformed to include a central charge. First of all, we address this issue
by considering the AdS extension \cite{BKsigma2}  of the Fayet-Sohnius hypermultiplet \cite{Sohnius}.

For further considerations it is useful to  introduce the U(1) generator
\bea 
\cJ \:= S^{kl} J_{kl} \ ,
\eea
which appears in (\ref{2.1}) and 
acts on the spinor covariant derivatives as follows
\bea 
[\cJ , \cD_\a^i ] = S^i{}_j \cD_\a^j \ , \qquad [\cJ , \cDb^\da_i] = - S^j{}_i \cDb^\da_j \ .
\eea
As noted in \cite{KT-M-AdS}, one can always choose
\be S^{\1\2} = 0 
\ee
by applying a rigid SU(2) rotation. This choice is very useful for reduction to $\cN=1$ AdS superspace 
and is assumed in what follows.
We denote the other components of ${ S}^{ij}$ as 
\bea
{S}^{\1\1}= { S}_{\2\2} = -\bar \m~, \qquad {S}^{\2\2} = {S}_{\1\1} = -\m~.
\label{S11S22}
\eea

\subsection{Fayet-Sohnius hypermultiplet}

Our presentation in this subsection follows  \cite{BKsigma2}. 
The Fayet-Sohnius hypermultiplet in AdS is described by a two-component superfield\footnote{Isospinor
indices are raised and lowed using antisymmetric tensors $\ve^{ij}$ and $\ve_{ij}$ normalized by 
$\ve^{\1\2} = \ve_{\2\1} =1$. The rules  are: 
$q^i = \ve^{ij} q_j$ and $q_i = \ve_{ij}q^j$. The conjugation property $ \overline{q_i} =\bar q^i $
implies $\overline{q^i}=-\bar q_i$.} 
$q_i$ and its conjugate 
$\bar q^i := \overline{q_i}$ subject to  the constraints
\bea 
\cD_\a^{(i} q^{j)} = \cDb_\da^{(i} q^{j)} = 0 \ .
\label{2.8}
\eea
Note that we do not assume a given action of $\cJ$ on $q_i$. 
Instead we rely on the constraints to determine its action. 
It follows from 
(\ref{AdSD})
that we may write
\be \cJ = \frac{1}{4} \{ \cDb_{\da \1} , \cDb^\da_\2 \} \ .
\ee
Using the constraints together with the covariant derivative algebra, one can show
\begin{subequations}
\begin{align}
\cJ q_{\1} &= - \frac{1}{4} (\cDb_\1)^2 q_\2 \ , 
\qquad \cJ \bar{q}_\1 = - \frac{1}{4} (\cDb_\1)^2 \bar{q}_\2~, \label{q1eq} \\
\cJ q_\2 &= \phantom{-}\frac{1}{4} (\cD^\1)^2 q_\1 \ , 
\qquad \cJ \bar{q}_\2 = \phantom{-}\frac{1}{4} (\cD^\1)^2 \bar{q_\1} \ .
\end{align}
\end{subequations}
Motivated by the fact that $q_\1$ 
and $\bar q_{\1}$
are  $\cN =1$ chiral superfields, 
\bea
\bar \cD_{\1}^\da q_{\1} =0~, 
\qquad \bar \cD_{\1}^\da \bar q_{\1} =0~,
\eea
we can rewrite \eqref{q1eq} as
\bea 
\cJ q_\1 + \mu q_\2 = - \frac{1}{4} [(\cDb_\1)^2 - 4 \mu] q_\2 ~, 
\qquad \cJ \bar q_\1 + \mu \bar q_\2 = - \frac{1}{4} [(\cDb_\1)^2 - 4 \mu] \bar q_\2 ~.
\eea
Then introducing $\dsJ$, the U(1) operator transforming $q_i$ as an isospinor,
\be 
\dsJ q_i := - S_i{}^j q_j ~, \qquad \dsJ \bar q_i := - S_i{}^j \bar q_j ~, 
\ee
we can write
\be 
\D q_\1 = - \frac{1}{4} [(\cDb_\1)^2 - 4 \mu] q_\2 \ , \qquad
\D \bar q_\1 = - \frac{1}{4} [(\cDb_\1)^2 - 4 \mu] \bar q_\2 ~.
\label{CCFS}
\ee
where we have introduced 
\bea
\D = \cJ - \dsJ~.
\eea 
Similarly we find
\be 
\D q_\2 = \frac{1}{4} [(\cD^\1)^2 - 4 \bar{\mu}] q_\1~, \qquad
 \D \bar q_\2 = \frac{1}{4} [(\cD^\1)^2 - 4 \bar{\mu}] \bar q_\1~.
\label{CCFS2}
\ee
Here $\D$ takes on the role of a central charge as it commutes with the covariant derivatives, 
\begin{align}\label{eq_SK8}
[\Delta, {\cD}_\alpha^i] = [\Delta, \bar{\cD}_{\da i}] = 0~.
\end{align} 
Thus  the constraints (\ref{2.8}) allow us
to specify the action of the generator $\cJ$ on the hypermultiplet 
as well as to separate 
a central charge, $\D$. We will use this procedure in the next subsection and for the VT multiplet in AdS superspace.

\subsection{Linear multiplet}

Since there exist interesting $\cN=2$ AdS supermultiplets with central charge, we have to construct 
a supersymmetric action principle to describe their dynamics. This can be achieved by 
generalizing the famous construction due to Sohnius \cite{Sohnius}.
The idea is to make a linear multiplet in ${\rm AdS^{4|8} } $ take on the role of a  superfield Lagrangian.

Following \cite{Sohnius,BS}, the linear multiplet 
is a real isotriplet superfield, $\cL^{ij}=\cL^{ji}$
and 
$\overline{\cL^{ij}}=\cL_{ij}$, 
 subject to the constraints
\bea 
\cD_\a^{(i} \cL^{jk)} = \bar \cD_\da^{(i} \cL^{jk)} = 0 \ .
\label{2.18}
\eea
We define $\cL^{ij}$  to transform under $\rm OSp(2|4)$, the isometry group of the $\cN=2$ 
AdS superspace, by the rule
\bea 
\d \cL^{ij}  = - {\bm \x} \cL^{ij}  - 2 {\bm \eps} \cJ \cL^{ij} ~, 
\label{2.19}
\eea
where the first-order operator $\bm \x$ and parameter $\bm \eps$
are given by eqs. (\ref{B.5}) and (\ref{B.3}) respectively. 
The U(1) generator $\cJ ={\mathbb J} +\D$ acts on the linear multiplet as 
\bea
\cJ \cL^{ij} = S^i{}_k \cL^{kj} + S^j{}_k \cL^{ki} + \D\cL^{ij}~.
\eea
In general, the linear multiplet has a non-zero central charge, $\D \cL^{ij} \neq 0$.\footnote{The case
 $\D \cL^{ij} =0$ corresponds to the  $\cN=2$ tensor multiplet.}
Indeed, the constraints (\ref{2.18}) imply that
\bea
\cJ \cL_{\1\1} = -\hf (\bar \cD_{\1} )^2 \cL_{\1\2} ~, \qquad
\cJ \cL_{\2\2} =  \hf ( \cD^{\1} )^2 \cL_{\1\2} ~.
\eea
These results imply the action of the central charge on $\cL_{\1\1}$ 
and $\cL_{\2\2}$:
\bea
\D \cL_{\1\1} = -\hf \big[ (\bar \cD_{\1} )^2-4\m \big] \cL_{\1\2} ~, \qquad
\D \cL_{\2\2} =  \hf \big[( \cD^{\1} )^2 -4\bar \m \big]\cL_{\1\2} ~.
\label{2.22}
\eea
In accordance with (\ref{2.18}), the superfield $\cL_{\1\1} $ is $\cN=1$ chiral, 
\bea
\bar \cD^\da_{\1} \cL_{\1\1} =0~.
\eea
The first equation in (\ref{2.22}) shows that $\D \cL_{\1\1}$ is also $\cN=1$ chiral, 
$  \bar \cD^\da_{\1} \D \cL_{\1\1} =0$.

\subsection{Supersymmetric action principle}  \label{SAP}

When dealing with  $\cN=2$ supersymmetric actions, it is convenient to use two types of reduction
with respect to the Grassmann variables: (i) reduction to $\cN=1$ superspace; and (ii) complete reduction. 
Given an $\cN = 2$ superfield $U(x, \theta_{\imath}, \bar{\theta}^{\imath})$, 
we define its $\cN = 1$ projection as
\be 
U| := U(x, \theta_{\imath}, \bar{\theta}^{\imath})|_{\theta_{\2} = \bar{\theta}^\2 = 0} \ ,
\ee
while its component projection is defined by 
\be 
U|\!| := U(x, \theta_{\imath} , \bar{\theta}^{\imath})|_{\theta_{\imath} = \bar{\theta}^{\imath} = 0} \ .
\ee

Associated with the linear multilplet, $\cL_{ij}$, is the following functional 
\bea 
S = - \frac{1}{12} \int \rd^4 x\, e\, \big( \cD^{ij} + \bar \cD^{ij} + 36 S^{ij} \big) \cL_{ij}|\!| \ ,
\qquad e^{-1} = \det (e_a{}^m) \ ,
\label{2.26}
\eea
where we have denoted $\cD^{ij} := \cD^{\a (i} \cD_\a^{j)} $
and $ \cDb_{ij} := \cDb_{\da (i} \cDb^{\da}_{ j)} $. It is assumed that $S$ is evaluated 
in a Wess-Zumino gauge of the form \cite{KT-M}
\bea
\cD_a|\!| &=& e_a+\o_a~,\qquad 
e_a=e_a{}^m(x)\pa_m~,\quad
\o_a=\hf\o_a{}^{bc}(x)M_{bc}~, 
\eea
with no U(1) connection being present in $\cD_a|\!|$.
The crucial property of the functional (\ref{2.26}) is that it 
turns out to be  invariant under arbitrary AdS isometry transformations 
(\ref{2.19}). This means that  (\ref{2.26}) can be used as supersymmetric action for theories in 
$\cN=2$ AdS superspace.

To show that the action  (\ref{2.26}) is indeed $\cN=2$  supersymmetric, 
we first reformulate it in  
$ \cN=1$ AdS superspace where one supersymmetry is manifestly realized.
Making use of the constraints (\ref{2.18}) we find
\bea 
- \frac{1}{12} (\cD^{ij}+ \cDb^{ij}  +36 S^{ij}) \cL_{ij} 
= - \frac{1}{4} [(\cD^\1)^2 - 12 \bar{\mu}] \cL_{\1\1} - \frac{1}{4} [(\cDb_\1)^2 - 12 \mu] \cL_{\2\2} \ .
\eea
Projecting this to $\cN=1$ AdS superspace gives
\bea 
- \frac{1}{12} (\cD^{ij}+ \cDb^{ij}  +36 S^{ij}) \cL_{ij} |
= - \frac{1}{4} (\nabla^2 - 12 \bar{\mu}) \cL_{\1\1}| - \frac{1}{4} (\bar \nabla^2 - 12 \mu) \cL_{\2\2}| \ .
\eea
The $\cN = 1$ AdS integration rule (see, e.g., \cite{KS94})
\bea 
\int \rd^4x \rd^4 \q \,\frac{E}{\mu} \, \cL_{\text{chiral}} = 
- \frac{1}{4} \int \rd^4 x \,e\, (\nabla^2 - 12 \bar{\mu} ) \cL_{\text{chiral}} |_{\q=\bar \q=0} ~,
\qquad \bar \nabla_\da \cL_{\text{chiral}} =0~~~~
\eea
can then be used to rewrite the action. One obtains
\bea 
S = \int \rd^4 x \rd^4\q \, \frac{E}{\mu} \, \cL_{\1\1} |
+ \int \rd^4 x \rd^4 \q \, \frac{E}{\bar{\mu}} \,\cL_{\2\2}| ~. \label{N1action}
\eea
In accordance with (\ref{2.22}),  this functional is invariant under central charge transformations 
\bea
\d \cL^{ij} = \z \D \cL^{ij}~, \qquad \z = \text{const}~.
\eea

Now we are finally prepared to prove invariance of  the action  (\ref{2.26}) under arbitrary $\cN=2 $
isometry transformations, eq. (\ref{2.19}). As shown in \cite{KT-M-AdS} and reviewed in \cite{BKsigma2}
and Appendix \ref{killing}, any $\cN=2$ isometry transformation induces two different transformations 
in $\cN=1$ AdS superspace which are: (i) an isometry transformation of $\cN=1$ AdS superspace 
which is generated by a Killing vector superfield 
$\x = \x^a \nabla_a + \x^\a \nabla_\a + \bar{\x}_\da \bar{\nabla}^\da$; 
and (ii) an extended supersymmetry transformation described by a real 
superfield parameter $\ve$ constrained as in eq. (\ref{B.12}).
The $\cN=2$ transformation law  (\ref{2.19}) implies that $\cL_{\1\1} |$ transforms as a scalar superfield 
under the $\cN=1$ AdS supergroup $\rm OSp(1|4)$, 
\bea
\d_\x  \cL_{\1\1} | =-\x \cL_{\1\1} |~.
\eea
The action (\ref{N1action}) is manifestly invariant under these transformations.

It remains to be shown that (\ref{N1action}) is also
invariant under the second supersymmetry. 
The second supersymmetry transformation of $\cL_{\1\1}|$ (see Appendix \ref{killing}) is
\be \d_\eps \cL_{\1\1} | 
= - \eps^\a \cD_\a^\2 \cL_{\1\1}| - \bar{\eps}_\da \cDb^\da_\2 \cL_{\1\1}| -2 \eps \cJ \cL_{\1\1}| ~.
\ee
Using the constraints on $\cL_{ij}$ and \eqref{epseq} we find
\begin{subequations}
\bea \d_\eps \cL_{\1\1}| &=& (\bar{\nabla}^2 - 4 \mu) (\eps \cL_{\1\2}) |\ , 
\eea
and similarly
\bea 
\d_\ve \cL_{\2\2}| &=& - (\nabla^2 - 4 \bar{\mu}) (\eps \cL_{\1\2})| \ .
\eea
\end{subequations}
These results imply that the action is supersymmetric as required.

\subsection{Hypermultiplet models}

To illustrate the supersymmetric action principle,  here we present examples involving the 
Fayet-Sohnius hypermultiplet. Given a linear multiplet, the corresponding action, \eqref{N1action}
 can be read directly from $\cL_{\1\1}|$. For the examples that we shall consider it is necessary to 
 first define the $\cN = 1$  projection of the 
Fayet-Sohnius hypermultiplet.

It follows from the constraints \eqref{2.8} that
\be \Phi_+ := q_{\1}| \ , \qquad \Phi_- = \bar{q}_{\1}| 
\ee
are covariantly chiral $\cN=1$ superfields,
\bea \bar{\nabla}_\da \Phi_+ = 0  \ , \qquad \bar{\nabla}_{\da} \Phi_- = 0 \ .
\eea
Furthermore performing the $\cN = 1$ projection of \eqref{CCFS} 
determines the action of the central charge on $\Phi_+$ and $\Phi_-$
\bea \D \Phi_+ = - \frac{1}{4} \big( \bar{\nabla}^2 - 4 \mu \big)  \Phib_- \ , \qquad \D \Phi_- 
= \frac{1}{4}  \big( \bar{\nabla}^2 - 4 \mu \big) \Phib_+ \ .
\eea
These results imply that 
\bea
(\D^2 + \Box_{\rm c}) \F_\pm =0~, 
\qquad \Box_{\rm c} := \frac{1}{16} \big( \bar{\nabla}^2 - 4  \mu \big)
\big( {\nabla}^2 - 4 \bar \mu \big)~,
\eea
with $\Box_{\rm c}$ being the covariantly chiral d'Alembertian. It is seen that the hypermultiplet becomes on-shell 
if $\D$ is set to be a constant  matrix.

A linear multiplet may be constructed using the hypermultiplet in a number of ways.  Firstly, we 
consider the linear multiplets
\bea (\cL_{\rm kin})_{ij} &=& \half \big( \bar{q}_{(i} \D q_{j)} - q_{(i} \D \bar{q}_{j)} \big) \ ,  
\label{2.37} \\
(\cL_{\rm der})_{ij} &=& \frac{\ri}{2} (\bar{q}_{(i} \D q_{j)} + q_{(i} \D \bar{q}_{j)}) \ .
\eea
However only the former leads to a kinetic term while the latter leads to a total derivative in components. 
A straightforward evaluation of $(\cL_{\rm kin})_{\1\1}|$ gives
\be (\cL_{\rm kin})_{\1\1}| = \half (\Phi_- \D \Phi_+ - \Phi_+ \D \Phi_-) 
= - \frac{1}{8} \big(\bar{\nabla}^2 - 4 \mu \big) (\Phi_+ \Phib_+ + \Phi_- \Phib_-) \ .
\ee
The corresponding action then reads
\bea S_{\rm kin} = \int \rd^4 x \rd^4 \q \,E \, (\Phi_+ \Phib_+ + \Phi_- \Phib_-) \ .
\eea

Another possible linear multiplet that we can consider is a bilinear in $q_i$ and $\bar{q}_i$ of the form
\bea 
(\cL_{\rm mass})_{ij} = \ri m \bar{q}_{(i} q_{j)} \ , \quad \bar{m} = m = \const \ ,
\label{2.41}
\eea
with corresponding action
\bea S_{\rm mass} = \ri m \int \rd^4 x \rd^4 \q \,\frac{E}{\mu} \,\Phi_+ \Phi_- + \cc
\eea

The specific feature of the actions generated by (\ref{2.37}) and (\ref{2.41}) is
invariance under U(1)
transformations
\be q_i \rightarrow \re^{\ri \varphi} q_i \ , \qquad \varphi \in \dsR \ .
\ee
This symmetry defines a charged hypermultiplet (when coupled to a Yang-Mills supermultiplet, 
such a hypermultiplet can transform in an arbitrary representation of the gauge group).
Without demanding this symmetry it is possible to construct an additional linear multiplet
\bea (\tilde{\cL}_{\rm mass})_{ij} = \half \cM q_i q_j + \half \cMb \bar{q}_i \bar{q}_j \ ,
\eea
with $\cM$ a complex mass parameter. The corresponding action is found to be
\bea \tilde{S}_{\rm mass} = \frac{1}{2} \int \rd^4x  \rd^4 \q\, \frac{E}{\mu} \,
\big( \cM  \Phi_+ \Phi_+ +  \bar \cM \F_-\F_- \big) ~+~\cc
\eea

\section{Vector-tensor multiplet}
\setcounter{equation}{0}
In the case of Poincar\'e supersymmetry, $\cN=2$ superfield techniques proved fruitful 
to obtain consistent formulations for the linear and nonlinear VT multiplets and their Chern-Simons 
couplings to $\cN=2$ vector multiplets \cite{GHH}--\cite{DIKST}.
This section examines the possibility of formulating a VT multiplet in $\cN = 2$ AdS superspace.

Here we derive consistent superfield constraints describing the nonlinear VT multiplet in AdS.
As proved in Section 1 with the use of $\cN=1$ superfield techniques, 
a linear VT multiplet does not exist in AdS. We present an 
alternative and more direct proof of this result using $\cN=2$  superfields.

\subsection{Consistent constraints}
Given some real superfield $\dsL$,  we make a general ansatz for the 
constraints
\begin{subequations}
\begin{align}
\cD^{ij} \dsL &= S^{ij} f(\dsL) + a \cD^i \dsL \cD^j \dsL+ b \cDb^i \dsL \cDb^j \dsL~,
 \label{AC1} \\
\cD_\a^{(i} \cDb_\da^{j)} \dsL &= c \cD_\a^{(i} \dsL \cDb_\da^{j)} \dsL \label{AC2} \ ,
\end{align}
\end{subequations}
for some function $f (\mathbb L )$ and parameters $a$, $b$ and $c$. The 
parameters are then fixed by consistency. There are two basic consistency requirements: \\
\indent (i) $\cD_\a^{(i} \cD^{jk)} \dsL = 0$; \\
\indent (ii) expressions for $\bar \cD_\da^{(i} \cD^{jk)} \dsL$ 
derived using \eqref{AC1} and \eqref{AC2} respectively coincide.

It is worth noting that as a consequence of the consistency conditions 
we cannot include additional terms in the ansatz 
without also including terms with two covariant derivatives of $\dsL$.
Imposing the consistency requirements yields two solutions: \\
\begin{subequations}
\begin{align} \cD^{ij} \dsL &= \frac{4}{\lambda} S^{ij} + \lambda \cD^i \dsL \cD^j \dsL ~,\\
\cD_\a^{(i} \cDb_\da^{j)} \dsL &= 0
\end{align}
\end{subequations}
and
\begin{subequations}
\begin{align}
 \cD^{ij} \dsL &= \frac{2}{\kappa} S^{ij} + 2 \kappa \cD^i \dsL \cD^j \dsL - \kappa \cDb^i \dsL \cDb^j \dsL~, \\
\cD_\a^{(i} \cDb_\da^{j)} \dsL &= \kappa \cD_\a^{(i} \dsL \cDb^{j)}_\da \dsL \ , 
\end{align}
\end{subequations}
where $\lambda$ is arbitrary and $\kappa$ is real. 
These solutions provide generalizations of the solutions found in \cite{DK,IS} and in 
an analogous way we can reject one of the solutions based on an additional consistency requirement, 
originating from the component structure of the multiplet. 
The superspace constraints give rise to differential constraints at the component level. 
We require that these component constraints can be solved for a gauge one-form and a gauge two-form. 
It turns out that only the second solution satisfies this deeper 
requirement and its component structure will be discussed in section \ref{components}. 
Furthermore, it is impossible to make the parameter $\kappa$ 
vanish and thus there exists no direct generalization of the linear VT multiplet in AdS, as expected.

It is useful to introduce a different superfield parameterization (compare with \cite{IS,DIKST})
\be L = \exp{(- \kappa \dsL)} \ ,
\ee
which takes the solution to the simpler  form
\begin{subequations}\label{constraints}
\begin{align} 
\cD^{ij} L &= - 2 S^{ij} L - \frac{1}{L} \cD^i L \cD^j L + \frac{1}{L} \cDb^i L \cDb^j L~, \\
\cD_\a^{(i} \cDb_\da^{j)} L &= 0 \ . 
\end{align}
\end{subequations}
We shall use this parameterization in the rest of the paper.

\subsection{Superfield Lagrangian}
It remains to find the corresponding linear multiplet which takes on the role of a Lagrangian density for the VT multiplet. To find a Lagrangian 
density for the VT multiplet corresponding to the constraints \eqref{constraints}, we try a general ansatz
\be \cL^{ij} = B(L) S^{ij} + A(L) \cD^i L \cD^j L + \bar{A}(L) \cDb^i L \cDb^j L \ ,
\ee
where $B(L)$ is an arbitrary real function and $A(L)$ is arbitrary. Imposing the constraints for a linear multiplet
\be 
\cD_\a^{(i} \cL^{jk)} = \cD_\da^{(i} \cL^{jk)} = 0
\ee
and using the constraints on $L$ leads to the  conditions
\be B'(L) = - 2 A(L) L \ , \ \ A'(L) = \frac{A(L)}{L} \ , \ \ \bar{A} = A \ .
\ee
These are solved by $A(L) = k L$ and $B(L) = -\frac{2 k}{3} L^3$, for some real $k$. 
Thus adopting a normalization for the Lagrangian density gives
\begin{align} 
\cL^{ij} &= \frac{1}{4} L (\cD^i L \cD^j L + \cDb^i L \cDb^j L - \frac{2}{3} L^2 S^{ij}) \non\\
&= \frac{1}{12} (\cD^{ij} + 4 S^{ij}) L^3 = \frac{1}{12} (\bar \cD^{ij} + 4 S^{ij}) L^3  \label{Ldensity}
\ ,
\end{align}
which generalizes the result in \cite{DK,IS}.

\subsection{Coupling to vector multiplets}

As an extension of the results in the previous two sections, here
we consider couplings of the VT multiplet with $\cN = 2$ super Yang-Mills fields.

The $\cN = 2$ super Yang-Mills multiplet in AdS superspace is described by a chiral field, 
$\cW$ obeying the constraints
\begin{align} \cDb_{\da i} \cW &= 0 \ , \non\\
(\cD^{ij} + 4 S^{ij}) \cW &= (\cDb^{ij} + 4 S^{ij}) \cWb \ .
\end{align}

We can choose the following consistent constraints in superspace, describing coupling of the VT multiplet to the Yang-Mills multiplet
\begin{subequations}\label{CS311}
\begin{align} \cD^{ij} L =& - 2 S^{ij} L - \frac{1}{L} \cD^i L \cD^j L + \frac{1}{L} \cDb^i L \cDb^j L \non  \\
&+ \frac{g}{2 L} \tr\big( (\cD^{ij} + 4 S^{ij}) \cF(\cW)  - (\cDb^{ij} + 4 S^{ij}) \cFb(\cWb) \big)~, \\
\cD_\a^{(i} \cDb_\da^{j)} L =& 0 \ ,
\end{align}
\end{subequations}
with $g$ a real coupling constant and $\cF(\cW)$ some holomorphic function. However it turns out that the corresponding components do not 
allow for an appropriate gauge two-form to be defined in general and we must choose the 
simplest nontrivial case 
\be 
\cF(\cW) = \cW^2 \ , \label{CS312}
\ee
which generates Chern-Simons terms at the component level (see section 5).

A corresponding Lagrangian density can then be constructed
\begin{align} 
\cL^{ij} =& \frac{1}{4} L (\cD^i L \cD^j L + \cDb^i L \cDb^j L - \frac{2}{3} L^2 S^{ij}) 
+ \frac{g}{8} \tr \Big(L (\cD^{ij} + 4 S^{ij}) \cW^2 + L (\cDb^{ij} + 4 S^{ij}) \cWb^2 \non\\
&- 2 (\cD^{ij} + 4 S^{ij}) (L \cW^2) - 2 (\cDb^{ij} + 4 S^{ij}) (L \cWb^2) \Big) \non\\
=& \frac{1}{12} (\cD^{ij} + 4 S^{ij}) (L^3 - 3 g L \tr (\cW^2) ) \non\\
&- \frac{g}{4} \tr \Big( (\cDb^{ij} + 4 S^{ij}) (L \cWb^2) - L (\cDb^{ij} + 4 S^{ij}) \cWb^2 \Big) \ ,
\end{align}
where reality of the last line follows from the constraints. This generalizes \eqref{Ldensity} to the case of Chern-Simons couplings.

\section{Formulation in N = 1 AdS superspace}
\setcounter{equation}{0}

Having derived the $\cN = 2$ constraints and Lagrangian density for the $\cN = 2$ VT multiplet in AdS superspace, 
it is natural to consider its formulation in terms of $\cN = 1$ superfields. 
This is especially apparent from the simplicity of the resulting action in $\cN=1$ superspace.
In this section we introduce $\cN = 1$ superfields 
in $\cN = 1$ AdS superspace describing the VT multiplet and its Chern-Simons coupling. 
We analyze the constraints obeyed by these superfields
and formulate the corresponding action principle. In particular, we demonstrate that the action principle of the 
VT multiplet possesses a rather simple structure in terms of cubic interactions.

\subsection{N  = 1 constraints}

We begin by reformulating the constraints in terms of $\cN =1$ superfields. 
Starting with the $\cN = 2$ super Yang-Mills 
multiplet we define the two independent $\cN = 1$ projections (see Appendix \ref{N1reduction}) as follows
\be 
\varphi := \cW| \ , \qquad \dsW_\a := \frac{\ri}{2} \cD_\a^{\underline{2}} \cW| \ .
\ee
A straightforward projection of the $\cN = 2$ constraints of $\cW$ give the following $\cN = 1$ constraints
\begin{subequations}
\begin{align} 
\bar{\nabla}_\da \dsW_\a &= 0~,  \\
\bar{\nabla}_\da \varphi &= 0 ~, \\ 
\nabla^\a \dsW_\a &= \bar{\nabla}_\da \bar{\dsW}^\da \ .
\end{align}
\end{subequations}
Thus $\varphi$ is a chiral superfield and $\dsW_\a $ describes a vector multiplet in $\cN = 1$ AdS superspace.

We are interested in the structure that the non-linear constraints on the VT multiplet possess. 
Firstly we note that the $\cN = 2$ VT superfield
contains two independent $\cN = 1$ projections defined as follows
\be
G := L| \ , \qquad W_\a := \half \cD_\a^{\underline{2}} L| \ .
\ee

The constraints for these $\cN = 1$ superfields follow from those on $L$, which are equivalent to
\begin{subequations} \label{Nconstraints}
\begin{align} (\cD^{\underline{1}})^2 L = &
- 2 S^{\underline{1}\underline{1}} - \frac{1}{L} \cD^{\underline{1}} L \cD^{\underline{1}} L 
+ \frac{1}{L} \cDb^{\underline{1}} L \cDb^{\underline{1}} L \non\\
& + \frac{g}{2L} \tr \big( (\cD^{\underline{1}})^2 + 4 S^{\underline{1} \underline{1}} ) \cW^2    \big) 
- \frac{g}{2L} \tr \big( (\cDb^{\underline{1}})^2 + 4 S^{\underline{1} \underline{1}} ) \cWb^2    \big)   \label{one} ~, \\
(\cD^{\underline{2}})^2 L =& - 2 S^{\underline{2} \underline{2}} L 
- \frac{1}{L} \cD^{\underline{2}} L \cD^{\underline{2}} L + \frac{1}{L} \cDb^{\underline{2}} L \cDb^{\underline{2}} L
\non \\
& + \frac{g}{2L} \tr \big( (\cD^{\underline{2}})^2 + 4 S^{\underline{2} \underline{2}} ) \cW^2    \big) 
- \frac{g}{2L} \tr \big( (\cDb^{\underline{2}})^2 + 4 S^{\underline{2} \underline{2}} ) \cWb^2    \big)    \label{two}  ~,\\
\cD^{(\underline{1}} \cD^{\underline{2})} L =& - \frac{1}{L} \cD^{\underline{1}} L \cD^{\underline{2}} L 
+ \frac{1}{L} \cDb^{\underline{1}} L \cDb^{\underline{2}} L + \frac{g}{2L} \tr \big( \cD^{\underline{1} \underline{2} }  \cW^2    \big) 
- \frac{g}{2L} \tr \big( \cDb^{\underline{1} \underline{2}}  \cWb^2    \big)   ~, \label{three}  \\
\cD_\a^{\underline{1}} \cDb_\da^{\underline{1}} L =& 0 \label{four} ~,\\
\cD_\a^{\underline{2}} \cDb^{\underline{2}}_\da L =& 0 \label{five} ~,\\
\cD_\a^{\underline{1}} \cDb^{\underline{2}}_\da L =& - \cD_\a^{\underline{2}} \cDb_\da^{\underline{1}} L \label{six} \ .
\end{align}
\end{subequations}
From \eqref{two} and \eqref{six} we can see that $\varphi$, $\dsW_\a$, $G$ and $W_\a$ form a basis for independent 
$\cN = 1$ projections. 
Namely, $\cN = 1$ projections formed out of two covariant derivatives of $L$ can always be written in 
terms of the $\cN=1$ superfields defined.

Analyzing the projection of the constraints \eqref{Nconstraints} lead to a number of $\cN = 1$ conditions. \eqref{one} 
and the constraint on $\cW$ gives a non-linear constraint for $G$
\begin{align} (\nabla^2 - 4 \bar{\mu}) G^2 &=  8 \Wb^2 + g \tr \big( (\nabla^2 - 4 \bar{\mu}) (\vf - \bar{\vf})^2 
+ 8 \bar{\dsW}^2 \big)   \ .
\end{align}
Then using \eqref{three} and its conjugate gives
\be \nabla^\a W_\a = \bar{\nabla}_\da \Wb^\da \ .
\ee
Finally, \eqref{four} and \eqref{five} imply chirality of $W_\a$
\be \bar{\nabla}_\da W_\a = 
0 \ .
\ee
Thus we have the $\cN = 1$ constraints for the VT multiplet with Chern-Simons terms
\begin{subequations} \label{N1C}
\begin{align} 
(\bar{\nabla}^2 - 4 \mu) \big( G^2 - g \,\tr (\vf-\bar \vf)^2 \big)&=  8 (W^2 -g\, \tr \dsW^2 )~, 
\\
\bar{\nabla}_\da W_\a &= 
0 ~,\\
\nabla^\a W_\a &= \bar{\nabla}_\da \Wb^\da \ .
\end{align}
\end{subequations}

The constraints \eqref{Nconstraints} also imply central charge transformations 
of the $\cN = 1$ superfields. Using \eqref{four} and \eqref{six} 
one derives
\begin{subequations} \label{N1CC}
\begin{align} 
G^2 \D G &= \hf \nabla^\a (G^2 W_\a) +\ri  g G \nabla^\a (\vf \dsW_\a) + \cc ~,\\
\D W_\a &= \frac{1}{8} (\bar \nabla^2 - 4 \mu) \nabla_\a G \ ,
\end{align}
\end{subequations}
where $\vf$ and $\dsW_\a$ are annihilated by the central charge. Furthermore from the above relations we deduce the supersymmetry
transformations (see appendix \ref{N1reduction})
\begin{subequations} \label{N1SUSY}
\begin{align} 
G^2 \d_\eps G &= - G^2 \nabla^\a \eps W_\a - \nabla^\a (\eps G^2 W_\a) - 2\ri g \eps G \nabla^\a (\vf \dsW_\a) + \cc ~,\\
\d_\eps W_\a &= - \frac{1}{4} (\bar \nabla^2 - 4 \mu) \nabla_\a (\eps G) ~,\\
\d_\eps \vf &= - 2 \eps^\a W_\a ~,\\
\d_\eps \dsW_\a &=  - \frac{\ri}{4} (\bar \nabla^2 - 4 \mu) \nabla_\a (\eps (\vf - \bar \vf)) \ .
\end{align}
\end{subequations}

\subsection{Supersymmetric action}

In section \ref{SAP} we presented the supersymmetric action
associated with the  linear multiplet. 
In particular we noted that the action 
can be written in terms of $\cN = 1$ projections. Here we make use of that result to derive the action rule 
for the VT multiplet in terms of  its $\cN = 1$ superfields.

Taking the $\cN = 1$ projection of $\cL_{\underline{1}\underline{1}}$ (and making use of the $\cN = 1$ constraints \eqref{N1C}) gives
\begin{align} 
\cL_{\underline{1}\underline{1}}| = \frac{1}{12} ( \bar{\nabla}^2 - 4 \mu) \Big(G^3 
+ 3 g G \,  \tr \big(\varphi^2  - 
\bar{\varphi}^2 \big) \Big) + 4\ri g \tr \big( W ^\a\dsW_\a \vf \big)  ~.
\end{align}
Putting this result in our action principle leads to
\begin{align} S &= - \frac{1}{3} \int  \rd^4 x \rd^4 \q \,E \, 
\Big(G^3 + 3 g G\,  \tr \big(\varphi^2  - \bar{\varphi}^2 \big) \Big) 
+ 4\ri g \int  \rd^4 x \rd^4 \q \,
\frac{E}{\mu}\, \tr \big( W^\a \dsW_\a \varphi \big) + {\rm c.c.} \non\\
&= - \frac{1}{3} \int  \rd^4 x \rd^4 \q \,E \,  G^3 
+ 4\ri  g \int  \rd^4 x \rd^4 \q \,\frac{E} {\mu} \, \tr \big( W^\a \dsW_\a \varphi \big) + {\rm c.c.} \non\\
&= - \frac{2}{3} \int  \rd^4 x \rd^4 \q \,E \,  G^3 
+ 4\ri g \int  \rd^4 x \rd^4 \q \,{E} \Big\{ \frac{1}{\mu} \,\tr \big( W \dsW \varphi \big) 
-
\frac{1}{\bar{\mu}} \, \tr \big( \Wb \bar{\dsW} \bar \vf \big) \Big\} ~,
\label{4.12}
\end{align}
where we have lifted part of the action from an integral over a chiral subspace to full superspace. 
As a check, one can show using \eqref{N1CC} 
and \eqref{N1SUSY} that the action is invariant under both central charge and supersymmetry transformations.

Turning off the Chern-Simons coupling reduces the action to 
\bea
S= - \frac{2}{3} \int  \rd^4 x \rd^4 \q \,E \,  G^3 ~.
\label{4.13}
\eea
Both actions (\ref{4.12}) and (\ref{4.13}) are cubic. The reason for this is that both theories 
are related to five-dimensional $\cN=1$ supersymmetric Chern-Simons theories \cite{KL}.

\section{Component results} \label{components}
\setcounter{equation}{0}

\allowdisplaybreaks

The superspace consistency conditions for the constraints of the VT multiplet 
do not guarantee the existence of a gauge one-form and 
a gauge two-form in its formulation. In order to verify their existence we must analyze the component fields.

We define the component fields of the external $\cN = 2$ Yang-Mills multiplet, $\cW$, as
\begin{align} 
w = \cW|\!|~,  \qquad \S_\a^i &= \cD_\a^i \cW| \!| ~, \qquad \bar \S^\da_i = \cDb^\da_i \cWb|\!| ~, \non\\
F_{\a\b} &= -\frac{1}{8} \cD_{\a\b} \cW |\!| \ , \qquad \Fb_{\da\db} = \overline{F_{\a\b}} ~,\non\\
X^{ij} &= (\cD^{ij} + 4 S^{ij}) \cW|\!| \ ,
\end{align}
and those of the VT multiplet as
\begin{align} 
l &= L|\!| \ , \qquad \lambda_\a^i = \cD_\a^i L|\!| \ , \qquad \bar{\lambda}_{\da i} = \cDb_{\da i} L|\!| \ , 
\qquad U = \Delta L|\!| ~,\non\\
V_{\a\da} &= - \frac{1}{4} [\cD_\a^i , \cDb_{\da i}] L|\!| = - \half \cD_\a^i \cDb_{\da i} L|\!| ~, \non\\
G_{\a\b} &= - \frac{\ri}{8} [\cD_\a^i , \cD_{\b i}] L|\!| 
= - \frac{\ri}{4} \cD_{\a\b} L|\!| \  , \qquad \Gb_{\da\db} = \overline{G_{\a\b}} \ . 
\end{align}
We note that the central charge transformations of $G_{ab}$ and $V_{a}$ are
\begin{subequations}
\begin{align} \Delta G_{ab} =& - 2 \cD_{[a} V_{b]} \ , \\
\D V_a =& - \frac{1}{l} V_a U - \frac{1}{4 l} \eps_{abcd} G^{bc} V^d - \frac{1}{l} \cD^b l G_{ab} - \cD^b G_{ab} \non\\
& + \frac{g}{2 l} \tr \big( 4 i \cD^b ((w - \bar w) F_{ab}) - 2 \eps_{abcd} \cD^b (w + \bar w) F^{cd} \non\\
&- \ri (\s_{ab})^{\a\b} \cD^b (\S_\a^i \S_{\b i}) - \ri (\tilde{\s}_{ab})^{\da\db} \cD^b (\bar{\S}_{\da i} \bar{\S}_\db^i) \big) \non\\
& + \frac{1}{2 l^2} (\s_a)^{\a\da} U \l_\a^i \bar{\l}_{\da i}  - \frac{\ri}{2l^2} (\s_a)^{\a\da} G_{\a\b} \l^{\b i} \bar{\l}_{\da i} + \frac{\ri}{2l^2} (\s_a)^{\a\da} \bar{G}_{\da\db} \l_\a^i \bar{\l}^\db_i \non\\
& - \frac{\ri}{2 l^2} (\s_{ab})^{\a\b} \cD^b l \l_\a^i \l_{\b i} - \frac{\ri}{2 l^2} (\tilde{\s}_{ab})_{\da\db} \cD^b l \bar{\l}^\da_i \bar{\l}^{\db i}  \non\\
& + \frac{1}{2 l^2} (\s_{ab})^{\a\b} V^b \l_\a^i \l_{\b i} - \frac{1}{2 l^2} (\tilde{\s}_{ab})_{\da\db} V^b \bar{\l}^\da_i \bar{\l}^{\db i}  \non\\
&+ \frac{1}{4 l^3} (\s_a)^{\a\da} \l_{\a i} \bar{\l}_{\da j} \l^{\b i} \l_\b^j - \frac{1}{4 l^3} (\s_a)^{\a\da} \l_{\a i} \bar{\l}_{\da j} \bar{\l}_\db^i \bar{\l}^{\db j} \non\\
& - \frac{g}{8 l^2} (\s_a)^{\a\da} \l_{\a i} \tr \big( -4 \ri \cD_{\b \da} \S^{\b i} (w - \bar w) - 4 \ri \S^{\b i} \cD_{\b \da} w + X^{ij} \bar{\S}_{\da j} \non\\
&- 8 \bar{F}_{\da\db} \bar{\S}^{\db i} - 4 S^{ij} \bar{\S}_{\da j} \bar w \big) \non\\
& - \frac{g}{8 l^2} (\s_a)^{\a\da} \bar{\l}_{\da i} \tr \big( 4 \ri \cD_{\a\db} \bar{\S}^{\db i} (w - \bar{w}) - 4 \ri \cD_{\a\db} \cWb \bar{\S}^{\db i} + X^{ij} \S_{\a j} \non\\
&- 8 F_{\a\b} \S^{\b i} - 4 S^{ij} \S_{\a j} w \big) \non\\
& - \frac{g}{4 l^3} (\s_a)^{\a\da} \l_{\a i} \bar{\l}_{\da j} \tr \big( X^{ij} (w - \bar{w}) - 2 S^{ij} (w^2 - \bar{w}^2) + \S^i \S^j - \bar{\S}^i \bar{\S}^j \big) \ .
\end{align}
\end{subequations}
Here $\cD_a$ denotes the space-time covariant derivative.\footnote{ Although this notation,
$\cD_a$,  coincides with that used earlier for the vector covariant derivative in $\rm AdS^{4|8}$, 
we hope no misunderstanding will occur.}
The superfield constraints lead to the following differential constraints on $F_{ab}$, 
$G_{ab}$ and $V_{a}$ 
\begin{subequations}
\begin{align} 
\cD_{[a} F_{bc]} =& 0 \ , \qquad \cD_{[a} G_{bc]} = 0 ~,\\
\cD^a H_a =& - \frac{1}{8} \eps^{abcd} G_{ab} G_{cd} - \frac{g}{2} \eps^{abcd} \tr \big( F_{ab} F_{cd}  \big) \non\\
& + \ri g \tr \cD^a \big( \cD_a w \bar{w} - \cD^a \bar{w} w + \ri (\s_a)_{\a\da} (\S^{\a i} \bar{\S}^\da_i) - \half \cD_a w^2 + \half \cD_a \bar{w}^2  \big) \ ,
\end{align}
\end{subequations}
where we define
\be H_a = l V_a + \half (\s_a)_{\a\da} \l^{\a i} \bar{\l}^\da_i \ .
\ee
Now, we can solve the constraints in terms of gauge one-forms $T_a$, $A_a$ and a two-form $B_{ab}$
\begin{subequations}
\begin{align}
F_{ab} =& 2 \cD_{[a} A_{b]} \ , \qquad G_{ab} = 2 \cD_{[a} T_{b]} \ , \\
H^a =&  \half \eps^{abcd} \Big( \cD_b B_{cd} - \frac{1}{4} T_b \cD_c T_d - g \tr \big( A_b \cD_c A_d \big)       \non\\
&{} \qquad + 2\ri g\, \tr \big( \cD_a w \bar{w} - \cD_a \bar{w} w 
+ \ri (\s_a)_{\a\da} \S^{\a i} \bar{\S}^\da_i - \half \cD_a w^2 + \half \cD_a \bar{w}^2 \big) \Big) \ .
\end{align}
\end{subequations}
This confirms the claim that the superfield constraints lead to a one-form and a two-form at the component level.

As a final note, we give the component action 
in the case where, for simplicity, the Chern-Simons coupling is turned off
\begin{align} S =& \int \rd^4x \,e \,
\Big( 
- \frac{1}{4} l G_{ab} G^{ab}   + \frac{1}{2 l} V^a V_a 
- \hf l \cD_a l \cD^a l 
+ \hf l U^2
+ \frac{1}{4} l^3 S^{ij} S_{ij} \non\\
& \qquad+
\frac{\ri}{2} G_{\a\b} \l^{\a i} \l^\b_i - 
\frac{\ri}{2}
\Gb_{\da\db} \bar{\l}^{\da}_i \bar{\l}^{\db i} 
- 
\frac{\ri}{2}
l \l^{\a i} \cD_{\a\da} \bar{\l}^\da_i + 
\frac{\ri}{2} 
l \cD_{\a\da} \l^{\a i} \bar{\l}^\da_i \non\\
& \qquad
+ \frac{1}{16 l} \l^i \l^j \l_i \l_j + \frac{1}{16 l} \bar{\l}_i \bar{\l}_j \bar{\l}^i \bar{\l}^j - \frac{3}{8 l} \l^i \l^j \bar{\l}_i \bar{\l}_j \Big) \ .
\end{align}

\section{Vector-tensor multiplet in  supergravity}\label{SUGRA}
\setcounter{equation}{0}

Having derived the appropriate constraints and Lagrangian density for 
the VT multiplet in AdS 
it is natural to 
look for 
an extension of our constructions to $\cN = 2$ supergravity.
We remind the reader that 
$\rm AdS^{4|8}$ is
a maximally symmetric geometry that originates within 
the superspace formulation of $\cN=2$ conformal supergravity developed in  \cite{KLRT-M08}
and reviewed in Appendix \ref{grimmspace}.
In the framework of supergravity, the central charge should be necessarily gauged. 
The  $\cN=2$ vector supermultiplet, which gauges the central charge, 
should be part of the so-called minimal multiplet of $\cN=2$ supergravity  \cite{BS}. 
The latter can be thought of as the $\cN=2$ Weyl multiplet  \cite{deWvHVP,BdeRdeW,deWLVP}  
coupled to the central charge vector multiplet.
Within the off-shell supergravity approach of \cite{KLRT-M08}, the action of any supergravity-matter system
should be invariant under super-Weyl transformations, see Appendix \ref{grimmspace}.
In particular, the VT multiplet constraints in supergravity should respect  super-Weyl invariance.

To describe the nonlinear VT multiplet, we
introduce a 
real scalar superfield  $L$ chosen (by analogy with the component approach of \cite{Claus3} 
and the rigid superspace construction of \cite{DIKST})
to be {\it inert} \footnote{We note that 
by making use of the central charge vector multiplet, $L$ can be redefined to a superfield 
$\hat L = L (\cZ \bar \cZ)^n$ 
with a  different super-Weyl transformation. }under the super-Weyl transformations, 
\bea
\d_\s L = 0~.
\label{6.3}
\eea
Making use of the central charge vector superfield, 
which is described by the covariantly chiral field strength 
$\cZ$ and its conjugate $\bar \cZ$,  we find consistent super-Weyl invariant constraints
\begin{subequations}\label{NLVT-SUGRA}
\begin{align}
\hf {\bm \cD}^{ij} L^2 &= \frac{\cZb}{\cZ} {\bm \cDb}^i L {\bm \cDb}^j L
- \frac{2}{\cZ} L{\bm  \cD}^{(i} \cZ {\bm \cD}^{j)} L 
- \frac{L^2}{2\cZ} ( {\bm \cD}^{ij} + 4 S^{ij}) \cZ ~,  \\
{\bm \cD}_\a^{(i} {\bm \cDb}_\da^{j)} L &= 0 \ ,
\end{align}
\end{subequations}
which generalize (\ref{constraints}). Here the gauge-covariant derivatives $\bm \cD_A $
are defined in eq. (\ref{D.7}).

To derive a linear multiplet $\cL^{ij}$, 
which governs the dynamics of the VT multiplet in supergravity, 
we have two requirements. Firstly, we require that the constraints
\be \bm \cD_\a^{(i} \cL^{jk)} = \bm \cDb_\da^{(i} \cL^{jk)} = 0 \ ,
\ee 
be satisfied. Secondly, we require  $\cL^{ij}$ to transform homogeneously under 
the super-Weyl transformations. 
Since the homogeneous super-Weyl transformation laws of covariant projective supermultiplets 
(to which $\cL^{ij}$ belongs)
are uniquely fixed \cite{KLRT-M08}, 
the super-Weyl transformation of $\cL^{ij}$ should be 
\be \d_\s \cL^{ij} = (\s + \bar{\s}) \cL^{ij} \ .
\ee
The corresponding linear multiplet satisfying the conditions given can then be constructed as
\be 
\cL^{ij} = \frac{1}{12} ({\bm \cD}^{ij} + 4 S^{ij}) (\cZ L^3)
= \frac{1}{12} (\bar{\bm \cD}^{ij} + 4 \bar S^{ij}) (\bar \cZ L^3)~.
\ee
These results generalize our formulation in AdS  and provide the first superspace
formulation of the nonlinear VT multiplet in $\cN = 2$ supergravity.

In accordance with the component analysis of Claus {\it et al.}
\cite{Claus3},  in $\cN=2$ supergravity a linear VT multiplet can be consistently defined 
in the presence of a second vector multiplet in addition to the central charge vector multiplet.
Within the superspace framework, such a supergravity-matter system can easily be constructed 
in conjunction with the rigid supersymmetric results of  \cite{DIKST}.
We  make use of an additional vector multiplet, 
described by the covariantly chiral field strength\footnote{The field strength $\cY$ obeys the constraints 
obtained from (\ref{D.10}) and (\ref{D.11}) by replacing $\cZ \to \cY$.  The super-Weyl transformation 
of $\cY$ is identical to that of $\cZ$, eq. (\ref{WeylTW}).}
$\cY$  and its conjugate $\bar \cY$,
to construct consistent super-Weyl invariant  constraints
\begin{subequations}\label{6.6}
\begin{align}
\bm \cD^{ij} L =&\phantom{-} \frac{2 \cYb}{\cZb \cY - \cZ \cYb} \big(\bm \cD^{(i} \cZ \bm \cD^{j)} L 
+ \bm \cDb^{(i} \cZb \bm \cDb^{j)} L + \hf L (\bm \cD^{ij} + 4 S^{ij}) \cZ \big) \non\\
&- \frac{2 \cZb}{\cZb \cY - \cZ \cYb} \big( \bm \cD^{(i} \cY \bm \cD^{j)} L 
+\bm \cDb^{(i} \cYb \bm \cDb^{j)} L + \hf L (\bm \cD^{ij} + 4 S^{ij}) \cY \big)~,  \\
\bm \cD_\a^{(i} \bm \cDb_\da^{j)}L =& 0 \ .
\end{align}
\end{subequations}
We note that although a pure linear VT multiplet does not exist in supergravity, the above is a consistent 
generalization of the constraints in the presence of the additional vector multiplet. 
In the flat superspace limit the constraints (\ref{6.6}) reduce to those given in \cite{DIKST}.
The corresponding 
Lagrangian density is given by
\begin{align} 
\cL^{ij} =& - \frac{\ri}{4} \big( \cY \bm \cD^i L \bm \cD^j L - \cYb \bm \cDb^i L \bm \cDb^j L\big)
+\frac{\ri}{8} \frac{\cY \cZb + \cZ \cYb}{\cZb \cY - \cZ \cYb} L^2 (\bm \cD^{ij} + 4 S^{ij}) \cY \non\\
&- \frac{\ri}{2} \frac{\cY \cYb L}{\cZb \cY - \cZ \cYb} \big(\bm \cD^{(i}\cZ \bm \cD^{j)} L 
+\bm \cDb^{(i}\cZb \bm \cDb^{j)} L + \hf L (\bm \cD^{ij} + 4 S^{ij}) \cZ \big) \non\\
&+\frac{\ri}{2} \frac{L}{\cZb \cY - \cZ \cYb} \big( \cZ \cYb \bm \cD^{(i} \cY \bm \cD^{j)} L 
+ \cZb \cY \bm \cDb^{(i} \cYb \bm \cDb^{j)} L \big) \ .
\end{align}
Its flat superspace limit coincides with that derived in \cite{DIKST}.

Although the constraints (\ref{NLVT-SUGRA}) and (\ref{6.6})
satisfy the basic consistency requirements, it is possible to formulate 
another consistency condition.\footnote{We are grateful to Daniel Butter for assistance 
with the derivation of this consistency condition.} 
It was noticed in \cite{DIKST} that after casting the constraints in terms of harmonic 
variables $u^{+i}$ and $u^-_i = \overline{u^{+i}}$ (normalized by $u^{+i}u^-_i =1$),  
one must demand $L$ to be independent of the harmonics. This leads to a non-trivial consistency 
requirement. Making use of the harmonics we generalize the condition to supergravity. Independence of 
harmonics leads to the condition
\be \bm \cD^{--} L = 0 \ ,
\ee
where $\bm \cD^{--} = u^{- i} \, \pa / \pa u^{+i} $  is one of the left-invariant vector fields on SU(2).
Applying successive gauged 
central charge covariant derivatives,
\be
\bm \cD_\a^{\pm} := u^\pm_i \bm \cD_\a^i~, \qquad 
\bm \cDb_\da^{\pm} :=  u^\pm_i \bm \cDb_\da^i~,
\ee 
to  the above condition leads to a number of relations. In particular, 
using the (anti-) commutation relations for the covariant derivatives,  
one derives
\begin{align}
0 =& \bm \cD^{+} \bm \cD^{+} \bm \cDb^{+} \bm\cDb^{+} \bm\cD^{--} L \non\\
=& \bm\cD^{--} \bm\cD^{+} \bm\cD^{+} \bm\cDb^{+} \bm\cDb^{+} L 
+ 8 \ri \bm\cD^{\a\da} \bm\cD_\a^{+} \bm\cDb^{+}_\da L 
- 2 \bm\cD^{-} \bm\cD^{+} \bm\cDb^{+} \bm\cDb^{+} L - 2 \bm\cDb^{-} \bm\cDb^{+} \bm\cD^{+} \bm\cD^{+} L \non\\
&- 4 \D \Big( (\bm\cD^{+})^2 (\cZ L) + (\bm\cDb^{+})^2 (\cZb L) + 2 L (S^{++} \cZ + \Sb^{++} \cZb) \non\\
&\qquad - \half L (\bm\cD^{+})^2 \cZ - \half L (\bm\cDb^{+})^2 \cZb \Big) \ .
\end{align}
This consistency condition places restrictions on the possible constraints for $L$. For instance, 
if we impose the constraint
\be \bm \cD_\a^{(i} \bm \cDb_\da^{j)}L = 0 \ ,
\ee
we have the condition
\begin{align}
0 =& \D \Big( (\bm\cD^{+})^2 (\cZ L) + (\bm\cDb^{+})^2 (\cZb L) + 2 L (S^{++} \cZ + \Sb^{++} \cZb) \non\\
& \qquad - \half L (\bm\cD^{+})^2 \cZ - \half L (\bm\cDb^{+})^2 \cZb \Big) ~,
\end{align}
which is equivalent to\footnote{One should keep in mind that the field strengths $\cZ$ and $\bar \cZ$
obey the Bianchi identity (\ref{D.11}).}
\begin{align}
0 =& \D \Big( (\bm\cD^{ij} +4S^{ij})(\cZ L) + (\bm\cDb^{ij} +4\bar S^{ij})(\cZb L) 
-  L (\bm\cD^{ij} +4S^{ij})\cZ 
\Big) ~.
\end{align}
We cannot impose the free constraint, $\bm\cD^{ij} L = 0$, without demanding 
annihilation of $L$ by the central charge $\D$, which would put $L$ on-shell.\footnote{It should 
pointed out that a constraint of the form $(\bm\cD^{ij} + \mu S^{ij})L = 0$, with $\mu$ a constant parameter, is not super-Weyl invariant and therefore it is not 
acceptable.}
Furthermore consistency for our supergravity 
constraints are guaranteed by the general super-Weyl invariant  identity 
\bea
0= ( \bm\cD^{ij} +4S^{ij}) (\cZ L) + ( \bm\cDb^{ij} +4\bar S^{ij})(\cZb L) 
-  L (\bm\cD^{ij} +4S^{ij})\cZ ~,
\eea
which holds for both the linear and nonlinear cases.

We also note that the super-Weyl freedom can be completely fixed by imposing the gauge 
\be
\cZ =1~.
\ee
This is known to restrict the torsion superfield $S^{ij}$ to be real, 
\bea
S^{ij}= \bar S^{ij} ~.
\label{7.2}
\eea

Given the linear multiplet, $\cL^{ij}$, the corresponding locally supersymmetric action is 
constructed in terms of the components of $\cL^{ij}$ and the central charge vector multiplet 
\cite{BS,deWvHVP2}. The same action takes a  compact form within the harmonic superspace 
approach to $\cN=2$ supergravity \cite{Galperin:1987em,Galperin:1987ek,GIOS}. 
As shown in  \cite{KT} the harmonic superspace action is 
\bea
S = \int \rd u \,\rd \z^{(-4)} \cV^{++} \hat{\cL}^{++}~.
\label{6.10} 
\eea
Here $\cV^{++}$ is the analytic gauge prepotential for the central charge vector multiplet, 
and $\hat{\cL}^{++}$ is obtained from 
\be
\cL^{++}:= \cL^{ij} u_i^+ u_j^+ 
\ee
by performing a  transformation  to the so called analytic frame.
The integration in (\ref{6.10}) is carried over the analytic subspace of harmonic superspace, 
see \cite{GIOS} for more details. The action (\ref{6.10}) is a natural generalization 
of the rigid supersymmetric action principle given  in \cite{DIKST}.

\section{Generalizations and further prospects}
\setcounter{equation}{0}

In this paper we have studied VT multiplets and their couplings to vector multiplets in AdS and, more generally, 
in $\cN=2$ supergravity within the superspace approach. In contrast to  the super-Poincar\'e case, 
the striking feature of AdS supersymmetry is non-existence of a free linear VT multiplet. 

Our results in section 6 provide the first superspace formulation of the nonlinear and the linear VT
multiplets in $\cN=2$ supergravity.  At the component level, 
a comprehensive study of the coupling of VT multiplets to $\cN=2$ supergravity was given in the past by Claus 
{\it et al.} \cite{Claus3}.
Comparing our results in section 6 with those 
derived  in \cite{Claus3}, one can see that the superfield constraints and Lagrangian densities 
are more compact than their component counterparts.
It was pointed out in  \cite{Claus3} that ``the complexity of our results clearly 
demonstrates the need for a suitable superspace formulation.'' 
Such a formulation has been developed in our paper. 

Using the locally supersymmetric constructions given in section 6, we can immediately 
derive new results in the case of AdS supersymmetry. It suffices
to `freeze' the background supergravity multiplet to a configuration describing the AdS geometry. 
This amounts to setting the torsion components $Y_{\a\b}$, $W_{\a\b} $ and $G_{\a \db }$ to vanish, 
\bea
Y_{\a\b}=0~, \qquad W_{\a\b} =0~,  \qquad G_{\a \db }=0
\eea
and also choosing  the remaining torsion $S^{ij}$ to be real, eq. (\ref{7.2}), 
and covariantly constant, $\cD_A S^{ij}=0$.
Upon such a reduction, the constraints (\ref{NLVT-SUGRA}) describe the nonlinear VT multiplet with gauged
central charge in AdS. We can further freeze the central charge vector multiplet 
to that having a constant field strength
\bea
\cZ = {\bm z} = \text{const}~, \qquad \bar{\bm z}=\bm z~.
\label{7.3}
\eea
Due to the Bianchi identity (\ref{D.11}) and the AdS condition (\ref{7.2}), 
the parameter $\bm z$ must be real.\footnote{The existence of a frozen vector multiplet, 
eq. (\ref{7.3}), in $\cN=2$ AdS superspace was proved in \cite{KT-M-AdS}.}
In the limit $\cZ \to {\bm z} =\bar{\bm z} ={\rm const}$,  the constraints (\ref{NLVT-SUGRA}) 
reduce to (\ref{constraints}).

Furthermore, upon freezing  the background supergravity multiplet to correspond to 
the AdS geometry, the constraints (\ref{6.6}) describe the linear VT multiplet in AdS in the presence of 
two vector multiplets one of which gauges the central charge. This formulation can be further
reduced to obtain two interesting special cases. First of all, in the AdS superspace we can freeze the $\cY$ vector multiplet to that having a constant field strength,  
\bea
\cY = {\bm y} = \text{const}~, \qquad \bar{\bm y}=\bm y~.
\label{7.3y}
\eea
This leads to the linear VT multiplet with gauge central charge in AdS
\begin{subequations}\label{7.4}
\begin{align}
(\bm \cD^{ij} +4S^{ij}) L =&\phantom{-} \frac{2 }{\cZb  - \cZ } \Big(\bm \cD^{(i} \cZ \bm \cD^{j)} L 
+ \bm \cDb^{(i} \cZb \bm \cDb^{j)} L + \hf L \bm \cD^{ij}  \cZ \Big)
~,  \\
\bm \cD_\a^{(i} \bm \cDb_\da^{j)}L =& 0 \ .
\end{align}
\end{subequations}
Secondly, we can further freeze the central charge vector multiplet 
to that having a constant field strength, eq. (\ref{7.3}). This leads to the 
linear VT coupled to a vector multiplet 
\begin{subequations}\label{7.5}
\begin{align}
( \cD^{ij} +4S^{ij})L =&
 \frac{2 }{  \cYb -\cY} \Big(  \cD^{(i} \cY  \cD^{j)} L 
+ \cDb^{(i} \cYb  \cDb^{j)} L + \hf L  \cD^{ij} \cY \Big) 
~,  \\
 \cD_\a^{(i}  \cDb_\da^{j)}L =& 0 \ .
\end{align}
\end{subequations}
The constraints (\ref{7.4}) and (\ref{7.5}) look formally identical to each other, but it should be 
kept in mind that the first set of constraints correspond to the case of gauge central charge. 
It is not possible to freeze
the remaining background vector multiplet in (\ref{7.4}) or (\ref{7.5}) 
to have a constant field strength since  the corresponding expectation value 
should be real in AdS, as emphasized in eqs. (\ref{7.3}) and (\ref{7.3y}),  
and hence the right hand side of (\ref{7.4}) or (\ref{7.5}) becomes singular when performing
a limit $\cZ \to \bm z$ or $\cY \to \bm y$.

Chern-Simons couplings of the VT multiplet, such as those described by the relations
(\ref{CS311}) and (\ref{CS312}), can be used as a tool to couple the VT multiplet 
to any number of $\cN=2$ tensor multiplets. This is achieved by making use of the techniques developed in 
\cite{Butter:2010jm} (see also \cite{deWS}).  It was shown in \cite{Butter:2010jm} how to generate 
a composite reduced chiral superfield,\footnote{The chiral field strengths of  Abelian vector multiplets 
are reduced chiral superfield.} $\mathbb W$, from a system of $n$ tensor multiplets 
described by their field strengths $G^{ij}_I$,  
with $I =1, \dots, n$,  
\bea
\cD^{(i}_\a G^{jk)}_I =  {\bar \cD}^{(i}_\da G^{jk)}_I = 0~.
\label{1.2}
\eea
The construction is as follows: 
\bea
{\mathbb W}  &=& \frac{1}{8\pi}  \oint_\g v^i {\rm d}v_i
 \Big( ({\bar \cD}^-)^2 +4 \bar{S}^{--}\Big) \O (G^{++}_I)~, \quad G^{++}_I:= v_i v_j G^{ij}_I  ~, 
 \quad v^i \in {\mathbb C}^2 \setminus \{0\}~,~~~
\label{1.55}
\eea
where  $\O(G^{++}_I)$  is a real homogeneous function 
of degree zero, $\O(c\,G^{++}_I)= \O(G^{++}_I)$, when $n>1$, and 
 $\O (G^{++} ) \propto \ln G^{++}$ in the case of a single tensor multiplet.
The integration in (\ref{1.55}) is carried over a closed contour $\g$ in ${\mathbb C}^2 \setminus \{0\}$. 
The right hand side of (\ref{1.55}) involves the second-order operator 
\bea
({\bar \cD}^-)^2 +4 \bar{S}^{--}:= \frac{u_i u_j}{(v,u)^2} \Big(\bar \cD^{ij} +4 \bar S^{ij}\Big)~, 
\qquad (v,u):=v^i u_i \ ,
\eea
which makes use of an isotwistor $u_i$ constrained by $(v,u) \neq 0$ 
and fixed along the integration contour $\g$; it can be shown that $\mathbb W$ is independent of $u_i$. 
As a simple example, we consider the case $n=1$ and   $\O (G^{++} ) = \ln G^{++}$ 
associated with the improved tensor multiplet \cite{deWPV,LR}.  For this choice,  eq. (\ref{1.55}) leads to 
\bea
\label{eq_imptensor}
\mathbb W =
     -\frac{G}{8} (\bar \cD_{ij} + 4 \bar S_{ij}) \Big(    \frac{G^{ij}}{G^2} \Big) ~, \qquad
     G^2 := \frac{1}{2} G^{ij} G_{ij}~.
\eea
Replacing $\cF (\cW) \to {\mathbb W}^2$ in (\ref{CS311}), with $\mathbb W$ given by (\ref{1.55}),
 yields a consistent higher-derivative 
coupling of the VT multiplet to $\cN=2$ tensor multiplets.

Our analysis of the supergravity-matter systems in section 6 treated the cases of 
the linear and nonlinear VT multiplets separately. We also ignored Chern-Simons couplings
to Yang-Mills vector multiplets. At the component level, Ref \cite{Claus3} provided a unified description 
of both the linear and nonlinear VT multiplets and their most general Chern-Simons couplings to 
vector multiplets. Using the results of our paper, it is possible to provide a superspace reformulation
and generalization
of the results in \cite{Claus3}. This will be discussed in a separate publication \cite{BN}.

Ten years ago, Theis \cite{Theis1,Theis2} constructed a new nonlinear VT multiplet
in Minkowski space. Interactions arise in this model as a consequence of gauging the central charge.
The latter is achieved by using the gauge one-form belonging to the VT multiplet, unlike the traditional 
approach of using a vector multiplet. It would be interesting to understand whether the construction 
of \cite{Theis1,Theis2} can be generalized to the case of AdS supersymmetry.

Recently, the supergravity results of \cite{Claus3} have been generalized in \cite{arXiv:1103.4813} 
to include $n_V$ vector and $n_T$ vector-tensor multiplets. As explained in \cite{arXiv:1103.4813}, 
their constructions could be obtained from standard $\cN=2$ supergravity coupled to $n_V+n_T$ 
vector multiplets by dualizing $n_T$ imaginary components of the $n_V+n_T$ complex scalar fields
parametrizing the special manifold. It would be interesting to understand how to obtain the results
of \cite{arXiv:1103.4813} in an off-shell superconformal setting. Clearly, the case $n_T=1$ 
is most interesting for string-theoretic applications \cite{deWKLL}.

In conclusion, let us summarize the main original results of this paper.
We developed the general superspace setting
for $\cN=2$ supersymmetric theories with central charge in AdS, including the supersymmetric  
action principle in $\cN=2$ AdS superspace.
We proved that $\cN=2$ AdS supersymmetry does not allow existence of a linear VT multiplet. 
For the nonlinear VT multiplet in AdS, we derived consistent superfield constraints in the presence 
of  any number of $\cN=2$ Yang-Mills vector multiplets, constructed the corresponding action 
and elaborated on the $\cN=1$ superfield and component descriptions 
of the  theory. 
For the superfield constraints and Lagrangians of \cite{DIKST}, 
 which describe the linear and the nonlinear VT multiplets with gauged central charge, 
we provided the consistent extensions to $\cN=2$ supergravity. 
We also constructed higher-derivative couplings 
of the VT multiplet to any number of $\cN=2$ tensor multiplets.
\\

\noindent
{\bf Acknowledgements:}\\
We are grateful to Daniel Butter for useful discussions and for reading the manuscript. 
The work  of SMK  is supported in part by the Australian Research Council.
The work of JN is supported by an Australian Postgraduate Award.

\appendix

\section{N = 1  AdS superspace} \label{N1ADS}
\setcounter{equation}{0}

In this appendix we collect salient facts about the geometry of  $\cN = 1$ AdS superspace
\cite{Keck,Zumino77,ISorin}, $\rm AdS^{4|4}$,  
and its isometries following \cite{BK}.

The geometry of 
$\rm AdS^{4|4}$
is determined by covariant derivatives,
\be
 \nabla_A = (\nabla_a, \nabla_\a, \bar{\nabla}^\da) 
= E_A{}^M \partial_M + \half \phi_A{}^{bc} M^{bc} \equiv E_A + \phi_A
\ee
obeying the following (anti-)commutation relations:
\begin{subequations}\label{A.2}
\begin{align} \{ \nabla_\a , \nabla_\b \} &= - 4 \bar{\mu} M_{\a\b} \ , \quad \{ \nabla_\a , \bar{\nabla}^\db \} 
= -2 \ri (\s^c)_\a{}^\db \nabla_c \equiv - 2 \ri \nabla_\a{}^\db \ , \\
[\nabla_a , \nabla_\b] &= - \frac{\ri}{2} \bar{\mu} (\s_a)_{\b\dg} \bar{\nabla}^\dg \ , 
\quad [\nabla_a , \nabla_b] = - |\mu|^2 M_{ab} \ ,
\end{align}
\end{subequations}
where $\mu$ is a complex non-vanishing parameter. Here the Lorentz generators with vector 
indices ($M_{ab} = - M_{ba}$) are related to those with spinor indices 
($M_{\a\b} = M_{\b\a}$) by the rules:
\be M_{ab} = (\s_{ab})^{\a\b} M_{\a\b} - (\tilde{\s}_{ab})^{\da\db} \Mb_{\da\db} \ , \quad M_{\a\b} = \half (\s^{ab})_{\a\b} M_{ab} \ , 
\quad \Mb_{\da\db} = - \half (\tilde{\s}^{ab})_{\da\db} M_{ab} \ . \non
\ee
The Lorentz generators act 
on the spinor covariant derivatives 
as
\be
[M_{\a\b}, \cD_\g] = \eps_{\g(\a} \nabla_{\b)} \ , 
\qquad [\Mb_{\da\db}, \bar{\nabla}_\dg] = \eps_{\dg (\da} \cDb_{\db )} \ ,
\ee
with $[M_{\a\b}, \cDb_\dg] = [\Mb_{\da\db} , \cD_\g] = 0$.

A real vector field, $\x^A = (\x^a, \x^\a, \x_\da)$, on $\rm AdS^{4|4}$ is called a Killing vector field if
\be 
[\x + \hf \l^{cd} M_{cd}, \nabla_A] = 0 \ , \quad \x 
:= \x^a \nabla_a + \x^\a \nabla_\a + \bar{\x}_\da \bar{\nabla}^\da \ ,
\label{A.4}
\ee
where $\l^{cd}$ is uniquely determined in terms of $\x^A$ and corresponds to some local Lorentz transformation. 
The master equation (\ref{A.4}) is  equivalent to
\begin{align}
\l_{\a\b} &= \nabla_\a \x_\b \ , \quad \nabla^\a \x_\a = 0 \ , 
\quad \frac{\ri}{2} \mu \x_{\a\da} + \bar{\nabla}_\da \x_\a = 0\ , \\
\nabla_{(\a} \x_{\b)\db} &= 0\ , \quad \bar{\nabla}^\db \x_{\a\db} + 8 \ri \x_\a = 0 \ ,
\end{align}
see \cite{BK} for a derivation. 
The AdS Killing vector fields generate the isometry group of the $\cN=1$ AdS superspace, 
$\rm OSp(1|4)$.
The infinitesimal isometry transformation associated with $\x^A$  
acts on a tensor superfield $U$ as follows
\be \d U = - \x U - \hf \l^{cd} M_{cd} U \ .
\ee

\section{N = 2 Killing vector fields} \label{killing}
\setcounter{equation}{0}

In this appendix we give a brief summary of the  Killing vector fields of $\cN=2$ AdS superspace, 
 $\rm AdS^{4|8}$,  
which were used in 
section \ref{N2ADS}. 
These objects were originally introduced in 
 \cite{KT-M-AdS} (see also  \cite{BKsigma2}).
A real vector field in $\cN = 2$ AdS superspace corresponding to the first-order operator 
\bea
{\bm \x} := {\bm \x}^{a}{\cD}_{ a} + {\bm \x}^\a_i { \cD}_\a^i + \bar {\bm \x}_\da^i \bar { \cD}^\da_i 
\label{B.5}
\eea
is said to be a Killing vector field if it obeys the master equation 
\be [ {\bm \x} + \half {\bm \l}^{cd} M_{cd}+ 2 {\bm \eps} \cJ, \cD_\a^i ] = 0~,  \qquad \cJ := S^{kl} J_{kl}
\ ,
\ee
 for uniquely determined parameters ${\bm \l}^{cd}$  and $\bm \eps$
 generating Lorentz and U(1) transformations respectively. 
 The explicit expressions for these parameters are
 \bea
 \bm\lambda_{ a b} = \cD_{ [a} \bm \xi_{ b]}~, \qquad
 {\bm \ve} = \frac{1}{8} { S}^{ij} { \cD}_{\a i} {\bm \x}^\a_j~,
 \label{B.3}
 \eea
 see \cite{BKsigma2,KT-M-AdS} for a derivation. 
 The Killing vector fields generate the isometry group of the $\cN=2$ AdS superspace, 
$\rm OSp(2|4)$.
If $U$ is  a tensor superfield in $\cN=2$ AdS superspace, its infinitesimal transformation associated with 
$\bm \x$ is  
\bea 
\d U = - {\bm \x} U - \half {\bm \l}^{cd} M_{cd} U - 2 {\bm \eps} \cJ U \ .
 \label{killertrans}
\eea

\section{N = 1 reduction} \label{N1reduction}
\setcounter{equation}{0}

Any supersymmetric field theory in $\cN=2$ AdS superspace,  $\rm AdS^{4|8}$,  
 can be reformulated in terms of 
superfields on $\cN=1$ AdS superspace \cite{KT-M-AdS},  $\rm AdS^{4|4}$. 
Such a reformulation proves to be useful 
for various applications. Here we give a summary of the $\cN=2 \to \cN=1$ reduction,  
more details can be found in \cite{BKsigma2, KT-M-AdS}.

Given a  tensor superfield $U(x, \theta_{\imath}, \bar{\theta}^{\imath})$
on  $\rm AdS^{4|8}$,  
its $\cN = 1$ projection is defined by
\be 
U| := U(x, \theta_{\imath}, \bar{\theta}^{\imath})|_{\theta_{\2} = \bar{\theta}^\2 = 0} 
\ee
in a {\it special coordinate system} specified below. 
Given a gauge-covariant operator of the form ${ \cD}_{A_1} \dots { \cD}_{A_n}$, 
its $\cN=1$ projection $\big({ \cD}_{A_1} \dots { \cD}_{A_n}\big)\big|$ is defined as follows:
\bea
 \Big( \big( { \cD}_{A_1} \dots { \cD}_{A_n}\big)\big| { U} \Big) \Big|
:=  \big({ \cD}_{A_1} \dots { \cD}_{A_n} U\big)\big|~, 
\eea
with $ U$ an arbitrary tensor superfield.
The required coordinate system is specified by 
\bea
{\cD}^\1_\a|=\nabla_\a~, \qquad { \cDB}^\da_\1|=\bar \nabla^\da~,
\eea
with $\nabla_{\rm A} = (\nabla_{ a} , \nabla_\a ,\bar \nabla^\da)$ the covariant derivatives
for $\rm AdS^{4|4}$ introduced in Appendix \ref{N1ADS}.
In such a coordinate system,
the operators ${ \cD}_\a^\1|$ and ${\cDB}_{\da \1}|$ do not involve any 
partial derivatives with respect to $\q_\2$ and ${\bar \q}^\2$, 
and therefore, for any positive integer $k$,  
it holds that $\big( { \cD}_{\hat{\a}_1} \cdots  { \cD}_{\hat{\a}_k}  U \big)\big|
= { \cD}_{\hat{\a}_1}| \cdots  {\cD}_{\hat{\a}_k}| U|$, 
where $ { \cD}_{\hat{\a}} :=( { \cD}_\a^\1, {\bar { \cD}}^\ad_\1)$ and $U$ is a tensor superfield.
We therefore obtain
\be
{ \cD}_{ a} | = \nabla_{ a}~.
\ee

The conceptual possibility to have a well-defined $\cN=2 \to \cN=1$ AdS superspace reduction
follows from the fact that the operators $(\cD_a,\cD_\a^\1, \cDb^\da_\1)$ 
form a closed algebra 
\begin{subequations}
\begin{align} \{ \cD_\a^\1 , \cD_\b^\1 \} &= 4 S^{\1\1} M_{\a\b} \ , \quad \{ \cD_\a^\1, \cDb^\db_\1\} 
= - 2 \ri \cD_\a{}^\db \ , \label{AdSspinor} \\
[\cD_a , \cD_\b^\1] &= \frac{\ri}{2} (\s_a)_{\b\dg} S^{\1\1} \cDb_\1^\dg \ , \quad  [\cD_a , \cD_b] = - S^2 M_{ab} \ ,
\end{align}
\end{subequations}
isomorphic to the covariant derivative algebra of $\cN = 1$ AdS 
superspace, eq.  (\ref{A.2}), with
\be \mu = - S_{\1\1} \ .
\ee

The isometries of  $\rm AdS^{4|8}$ are generated by the $\cN=2$ Killing vector fields.
Given such a Killing vector,  $\bm \x$, it induces two different transformations in
 $\rm AdS^{4|4}$ defined in terms of its $\cN=1$ projection
\be {\bm \x}| = \x + \x_\2^\a| \cD_\a^\2| + \bar{\x}_\da^\2| \cDb^\da_\2| \equiv \x + \eps^\a \cD_\a^\2| + \bar{\eps}_\da \cDb^\da_\2| \ .
\ee
Here $\x = \x^a \nabla_a + \x^\a \nabla_\a + \bar{\x}_\da \bar{\nabla}^\da$ 
proves to be  a Killing vector of the $\cN = 1$ AdS superspace. 
It can be shown that 
\bea
\eps_\a &= \nabla_\a \eps ~, \qquad \eps := {\bm \eps}|~.
\eea
The real parameter $\eps$ satisfies the constraints \cite{GKS}
\begin{subequations} \label{B.12}
\begin{align}  
(\bar{\nabla}^2 - 4 \mu) \eps &= (\nabla^2 - 4 \bar{\mu}) \eps = 0 \ , \label{epseq} \\
\nabla_\a \bar{\nabla}_\da \eps &= \bar{\nabla}_\da \nabla_\a \eps = 0 \ .
\end{align}
\end{subequations}
The parameters $\x$ and $\ve$ generate two different transformations. 
The former generates an isometry transformation of $\rm AdS^{4|4}$ acting on $U|$ by 
\be 
\d_\x U| = - \x U| - \half \l^{cd} M_{cd} U| ~.
\ee
The latter generates $U(1)$ and second supersymmetry transformations, 
\be \d_\ve U| = -  \eps^\a (\cD_\a^\2 U)| - \bar{\eps}_\da (\cDb^\da_\2 U)| 
 - 2 \eps \cJ U| \ .
\ee

\section{N = 2 conformal supergravity} \label{grimmspace}
\setcounter{equation}{0}
\allowdisplaybreaks

This appendix contains a summary of the  superspace formulation for $\cN=2$ conformal supergravity
developed in \cite{KLRT-M08}.  The formulation is based on the curved superspace geometry 
introduced by Grimm \cite{Grimm}. There exists a more general superspace formulation 
for $\cN=2$ conformal supergravity developed by Howe \cite{Howe}.\footnote{Howe's formulation 
\cite{Howe} is a gauged fixed version of the superspace formulation  
for $\cN=2$ conformal supergravity developed by Butter \cite{Butter:2011sr}.}
The precise relationship between these two formulations is spelled out in \cite{KLRT-M09}.
The results in this section are presented almost identically to \cite{KLRT-M08}.

Conformal supergravity can be realized in a curved 4D $\cN = 2$ superspace, $\cM^{4|8}$ parametrized by local coordinates $z^M = (x^m, \theta^\mu_\imath, \bar{\theta}_{\dot{\mu}}^\imath 
= (\theta_{\mu \imath})^* )$, where $m = 0, 1, ... \ , 3$, $\mu = 1, 2$. $\dot{\mu} = 1, 2$ and $\imath = \1, \2$.
The structure group is chosen to be $\rm SL(2, \dsC) \times SU(2)$, 
and the covariant derivatives $\cD_A = (\cD_a, \cD_\a^i, \cDb^\da_i)$ have the form
\begin{align} \cD_A &= E_A + \Phi_A{}^{kl} J_{kl}+ \Omega_A{}^{bc} M_{bc} \non \\
		  &= E_A + \Phi\indices{_A^{kl}} J_{kl}+ \Omega\indices{_A^{\b\g}} M_{\b\g} 
		  + \bar{\Omega}\indices{_A^{\dot{\b}\dot{\g}}} \Mb_{\dot{\b}\dot{\g}}~.
\end{align}
Here $M_{cd}$ and $J_{kl}$ are the generators of the Lorentz and SU(2) groups repectively, 
and $\Omega_A{}^{bc}$ and $\Phi_A{}^{kl}(z)$ the corresponding connections. 
The action of the generators on the covariant derivatives are defined as:
\begin{align} [M_{\a\b}, \cD_\g^i] &= \eps_{\g (\a} \cD_{\b)}^i \ , \qquad [\Mb_{\da\db} , \cDb^i_\dg] = \eps_{\dg (\da} \cDb^i_{\db) i} ~,\\
[J_{kl}, \cD_\a^i] &= - \d^i_{(k} \cD_{\a l)} \ , \qquad [J_{kl}, \cDb^\da_i] = - \eps_{i ( k} \cDb^\da_{l)} \ .
\end{align}
The covariant derivatives satisfy the (anti-)commutation relations \cite{Grimm}:
\begin{subequations}
\begin{align} \{ \cD_\a^i, \cD_\b^j \} =& 4 S^{ij} M_{\a\b} + 2\eps^{ij} \eps_{\a\b}Y^{\g\d} M_{\g\d} + 2\eps^{ij}\eps_{\a\b}\bar{W}^{\dg\dd}\bar{M}_{\dg \dd} \non \\
&+2\eps^{ij}\eps_{\a\b}S^{kl}J_{kl}+4Y_{\a\b}J^{ij}~, \\
\{\cD_\a^i , \cDb^\db_j \} =& - 2 i \d^i_j \cD_\a{}^\db +4 \d^i_j G^{\b\db} M_{\a\b}+4 \d^i_j G_{\a\da} \bar{M}^{\da\db}+8G_\a{}^\db J^i{}_j ~,\\
[\cD_{\a \da}, \cD_\b^j] =& 2 \ri \eps_{\a ( \b} G_{\g ) \da} \cD^{\g j} - \ri (\eps^{jk} \eps_{\da \dg} Y_{\a \b} + \eps^{jk} \eps_{\a \b} \Wb_{\da \dg} + \eps_{\a \b} \eps_{\da \dg} S^{jk} ) \cDb^\dg_k \non\\
&+ \frac{\ri}{2} (2 \eps_{\b ( \a} \cDb_\da^j Y_{\g ) \d} + \eps_{\a \g} \eps_{\b \d} \cDb^{\dg j} \Wb_{\da \dg} + \eps_{\a \g} \eps_{\b \d} \cDb_{\da l} S^{jl}) M^{\g \d} \non\\
&- \frac{\ri}{2} (\eps_{\da \dg} \cDb_\dd^j Y_{\a \b} + 2 \eps_{\a \b} \cDb_{(\da}^j \Wb_{\dg)\dd}+\frac{1}{3}\eps_{\a \b} \eps_{\da \dg} \cDb_{\dd l} S^{jl} ) \Mb^{\dg \dd} \non\\
&- \frac{\ri}{2} (2 \eps^{jk} \cDb_{\da}^l Y_{\a\b} + 2 \eps_{\a \b} \eps^{jk} \cDb^{\dd l}\Wb_{\da\dd}+\eps_{\a\b}\cDb_\da^jS^{kl}) J_{kl} \ .
\end{align}
\end{subequations}

It was shown in \cite{KLRT-M08} that the superspace geometry introduced is invariant 
under super-Weyl transformations:\begin{subequations} 
\label{WeylTD}
\begin{align} \d_\s \cD_\a^i =& \half \bar{\s} \cD_\a^i + (\cD^{\g i} \s) M_{\g\a} - (\cD_{\a k} \s) J^{ki} ~, \\
\d_\s \cD_a =& \half (\s + \bar{\s}) \cD_a + \frac{\ri}{4} (\s_a)^\a{}_{\db} (\cD_\a^k \s) \cDb^\db_k 
+ \frac{\ri}{4} (\s_a)^\a{}_\db (\cDb^\db_k \bar{\s}) \cD_\a^k \non \\
&  \qquad- \half \big(\cD^b(\s -\bar{\s})\big) M_{ab} \ , 
\end{align}
\end{subequations}
where $\s$ is an arbitrary covariantly chiral superfield, $\bar \cD^\da_i \s =0$.
The torsion components then transform as:
\begin{subequations} \label{WeylTTorsion}
\begin{align} \d_{\s} S^{ij} &= \bar{\s} S^{ij} - \frac{1}{4} \cD^{\g(i} \cD_{\g}^{j)} \s \ , \\
\d_\s Y_{\a\b} &= \bar{\s} Y_{\a\b} - \frac{1}{4} \cD_{(\a}^k \cD_{\b) k} \s \ , \\
\d_\s W_{\a\b} &= \s W_{\a\b} \ , \\
\d_\s G_{\a\db} &= \half (\s +\bar{\s}) G_{\a\db} - \frac{\ri}{4} \cD_{\a\db} (\s - \bar{\s}) \ .
\end{align}
\end{subequations}

Following \cite{KLRT-M09}, it is possible to incorporate a gauged central charge, $\D$, into 
the supergeometry by modifying the covariant derivatives to contain a central charge gauge connection, $V_A$:
\be 
\cD_A \rightarrow \bm \cD_A := \cD_A + V_A \D ~, \qquad \D V_A =0~. \label{D.7}
\ee
The corresponding algebra of gauge-covariant derivatives then becomes:
\be [ \bm \cD_A , \bm \cD_B \} = T_{AB}{}^C {\bm \cD}_C 
+ \half R_{AB}{}^{cd} M_{cd} + R_{AB}{}^{kl} J_{kl} + F_{AB} \D \ ,
\ee
where the gauge-invariant field strength, $F_{AB}$ is subject to covariant constraints and the torsion and curvature remain the same as the case without central charge. The components of $F_{AB}$ are:
\begin{subequations}
\begin{align} F_\a^i{}_\b^j =& - 2 \eps_{\a\b} \eps^{ij} \cZb \ , 
\qquad F^\da_i{}^\db_j = 2 \eps^{\da\db} \eps_{ij} \cZ \ , \qquad F_\a^i{}^\db_j = 0 ~,\\
F_a{}_\b^j =& \frac{\ri}{2} (\s_a)_\b{}^\dg \cDb_\dg^j \cZb \ , \qquad 
F_a{}^\db_j = - \frac{\ri}{2} (\s_a)_\g{}^\db  \cD^\g_j \cZ \ , \\
F_{ab} =& - \frac{1}{8} (\s_{ab})_{\a\b}  \cD^{\a k}  \cD_k^\b \cZ 
+ \frac{1}{8} (\tilde{\s}_{ab})_{\da\db}  \cDb^{\da}_k  \cDb^{\db k} \cZb \non\\
&+\half \big( (\tilde{\s}_{ab})_{\da\db} \Wb^{\da\db} - (\s_{ab})_{\a\b} Y^{\a\b} \big) \cZ 
- \half \big( (\s_{ab})_{\a\b} W^{\a\b} - (\tilde{\s})_{\da\db} \Yb^{\da\db} \big) \cZb \ ,
\end{align}
\end{subequations}
where $\cZ$ is a covariantly chiral superfield,
\be \cDb_{\da}^i \cZ = 0 \ ,
\label{D.10}
\ee
obeying the Bianchi identity
\be \big( \cD^{\g (i}  \cD_\g^{j)} + 4 S^{ij} \big) \cZ 
= \big(  \cDb_\dg^{(i}  \cDb^{\dg j)} + 4 \Sb^{ij} \big) \cZb \ .
\label{D.11}
\ee
To be consistent with the central charge interpretation, the field strength $\cZ$ should be nowhere vanishing, 
$\cZ \neq 0$. 
Super-Weyl transformations can then be seen to remain the same as in \eqref{WeylTD} and \eqref{WeylTTorsion} with
\be \d_\s \cZ = \s \cZ \ . \label{WeylTW}
\ee

\begin{footnotesize}

\end{footnotesize}

\end{document}